\documentclass[journal]{IEEEtran}
\usepackage{cite}
\ifCLASSINFOpdf
\usepackage[pdftex]{graphicx}
\else
\usepackage[dvips]{graphicx}
\fi
\usepackage{array}
\usepackage{amsmath}
\usepackage{amssymb}
\usepackage{amsthm}
\usepackage{titlesec}
\usepackage{graphicx,epstopdf}
\usepackage{tabularx}
\usepackage[none]{hyphenat}

\usepackage{array}
\usepackage{flushend}
\usepackage{multirow}
\titlespacing*{\section}{0pt}{1.0ex}{0.5ex}    
\titlespacing*{\subsection}{0pt}{1.2ex}{0.1ex}
\usepackage{draftwatermark}
\SetWatermarkText{International Journal of Electrical Power and Energy Systems
 (an Elsevier Journal)}
\SetWatermarkColor[gray]{0.1}
\SetWatermarkFontSize{0.75cm}
\SetWatermarkAngle{90}
\SetWatermarkHorCenter{20.5cm}

\begin{document}
\title{Deep Ensemble Learning-based Approach to Real-time Power System State Estimation}

\author{Narayan Bhusal*, Raj Mani Shukla**,  Mukesh Gautam*, Mohammed~Benidris*, and Shamik Sengupta**, \\ *Department of Electrical and Biomedical Engineering, University of Nevada-Reno, Reno, NV 89557, USA \\ **Department of Computer Science and Engineering, University of Nevada, Reno, Reno, NV 89557,  \vspace{-1ex}} 

\maketitle
\thispagestyle{plain}
\pagestyle{plain}
\begin{abstract}
Power system state estimation (PSSE) is commonly formulated as weighted least-square (WLS) algorithm and solved using iterative methods such as Gauss-Newton methods. However, iterative methods have become more sensitive to system operating conditions than ever before due to the deployment of intermittent renewable energy sources, low carbon technologies (e.g., electric vehicles), and demand response programs. Appropriate PSSE approaches are required to avoid pitfalls of the WLS-based PSSE computations for accurate prediction of operating conditions. This paper proposes a data-driven real-time PSSE using a deep ensemble learning algorithm. In the proposed approach, the ensemble learning setup is formulated with dense residual neural networks as base-learners and multivariate-linear regressor as meta-learner. Historical measurements and states are utilised to train and test the model. The trained model can be used in real-time to estimate power system states (voltage magnitudes and phase angles) using real-time measurements. Most of current data-driven PSSE methods assume the availability of a complete set of measurements, which may not be the case in real power system data-acquisition. This paper adopts multivariate linear regression to forecast system states for instants of missing measurements to assist the proposed PSSE technique. Case studies are performed on various IEEE standard benchmark systems to validate the proposed approach. The results show that the proposed approach outperforms existing data-driven PSSE methods techniques. 

\end{abstract}



\begin{IEEEkeywords}
Deep ensemble learning, multivariate linear regression, power system state estimation, residual neural net.
\end{IEEEkeywords}

\section*{Nomenclature}
\begin{tabbing}
AAAAAA \= AAAAAAAAAAAAAAAAAAAAAAAAAAAAAAAA \kill 

$z$\>measurement vector \\
$x$\>state vector\\
$e$\>measurement residual vector \\
$h$\> nonlinear function, relating state vector to \\ \> measurement vector \\
$H$\>Jacobian matrix \\
$G$\>gain matrix \\
$\zeta$\>set of all buses \\
$n$\> total number of buses in a power network \\
$\zeta_t^v$ and $\zeta_t^{\delta}$\>set of buses at which voltage magnitude and \\ \> phase angle measurements are available at any \\ \> instant $t$, respectively\\
$V_t^i$\>voltage magnitude measurement at bus $i$ at any \\ \>instant $t$ \\
$\delta_t^i$\>phase angle measurement at bus $i$ at any instant $t$ \\
$\zeta_t^p$ and $\zeta_t^{q}$\>set of buses at which real and reactive power \\ \> measurements are available at any instant $t$, \\ \> respectively\\
$P_t^i$\>real power injection measurement at bus $i$ at any \\ \> instant $t$ \\
$Q_t^i$\>reactive power injection measurement at bus $i$ at \\ \> any instant $t$ \\
$\pounds$\>total number of branches in a power network \\
$\pounds_t^p$ and $\pounds_t^q$\>set of branches at which real and reactive power \\ \> flow  measurements are available at any instant $t$, \\ \> respectively\\ 
$P_t^{ii'}$\>real power flow measurement from line $i$ to $i'$ at \\ \> any instant $t$ \\
$Q_t^{ii'}$\>reactive power flow measurement from line $i$ to $i'$ \\ \> at any instant $t$

\end{tabbing}

\section{Introduction}\label{}
 Power System State Estimation (PSSE) is used to provide real-time database for control and monitoring systems of power grids and to assist system operators in making well-informed remedial action decisions in case of contingencies. PSSE techniques use power system measurements like line flows, nodal voltages (magnitude and phase angle), and nodal injections (obtained from supervisory control and data acquisition (SCADA) system), to estimate power system states such as voltage magnitudes and phase angles at system nodes. PSSE is a non-convex problem generally formulated based on weighted least square (WLS) methods and solved using iterative methods such as Gauss-Newton methods. However, these methods are sensitive to system operating conditions and uncertainties. Also, the increase in the deployment of intermittent renewable energy sources (e.g., photovoltaic and wind power generation), power system-dependent low carbon technologies (e.g., electric vehicles), and load demand in modern power grids has led to frequent and sizeable voltage fluctuations. Furthermore, disparate cyber-attacks and natural events exacerbate the operation and control of power systems \cite{8966351, 9183679}. The aforementioned unpredictable  behaviour of power grids makes conventional solutions for PSSE, like WLS-based methods, computationally expensive and sub-optimal. Therefore, it is important to develop computationally efficient and technically feasible alternative solutions for power system state estimation that address the above-mentioned uncertainties.
 
Several model-based methods have been proposed in the literature to solve the PSSE problem using a number of statistical criteria. Most commonly used criteria are (a) maximum likelihood criterion---it maximises the probability of the estimated states being equal to true values of the states; (b) WLS criterion---it minimises the sum of the squares of the weighted errors of actual measurements and the estimated measurements; (c) minimum variance criterion---it minimises the expected value of the sum of squares of errors between the true states and estimated states; (d) least absolute value criterion---it minimises the sum of the absolute values of the deviations between estimated and actual measurements \cite{WoodWoolenberg658, 260812, 7742899, 6213712, 8632764, FAN2013168, WU199080}. Other criteria include minimum mean squared estimator \cite{8279475}, Schweppe Huber generalised M-estimator, and least-median and least-trimmed square estimator \cite{496203}. However, model-based approaches are sensitive to initialisation, require several iterations, are computationally intensive, and produce sub-optimal performance, specifically with newly emerging uncertainties and growing system dynamics. 

The development of various machine learning (ML) approaches has led to the use of big data in potpourri of complex  power system problems, and PSSE is not an exception \cite{8752583, AKHAVANHEJAZI201891, 10.1007/978-981-15-0135-7_38}.  Data-driven PSSE approaches provide great flexibility and scalability. They also  have the capability to improve the run-time efficiency and accuracy of conventional state estimation approaches \cite{8693907}. In \cite{8962079, 8693907, bretas2020distribution}, a hybrid of data-driven and statistical criteria (e.g., WLS and least absolute value) have been used to estimate power system states. In the hybrid approach, ML models (neural network (NN) in \cite{8962079} and long-short-term-memory (LSTM) in \cite{8693907}) have been used as a surrogate model to map available measurements or historical states in the neighbourhood of the true latent states. These approximate states have been used as initialisation for model-based criterion. Numerous model-free data-driven approaches have also been proposed in literature. These approaches use historical similar measurements or simulated measurements and their states for the training, validation, and testing of the ML models. The trained models have been used for real-time state estimation. The ML approaches such as k-nearest neighbour in \cite{7378513}, LSTM in \cite{ADENERGIES}, physics-inspired unrolled deep neural network (DNN) \cite{8754766}, auto-associated neural network \cite{6814454},  physics-aware NN \cite{9072507}, deep recurrent neural network \cite{8683139},  and deep generative adversary network \cite{8909752}, to name a few, have been used for PSSE. A comparison between the proposed approaches with some of these approaches is provided in the case studies.

The aforementioned data-driven approaches provide rich literature on PSSE and contribute toward the development of resilient future power grids. However, shallow neural network models suffer from scalability and computational inefficiencies \cite{SPRINGERs11633}. Furthermore, due to the stochastic nature of DNN, they are sensitive to a specific set of training data, which in turn results in different predictions---different sets of weights may be obtained every single time they are trained. In other words, the stochastic nature of DNN poses high variance and makes the development of final prediction models difficult.

This paper proposes a data-driven real-time PSSE model using deep ensemble learning method. Actual historical data (obtained from SCADA) and simulation-derived data (sampled snapshots using MATPOWER) are utilised to train several parallel dense Residual Neural Networks (ResNetD). The ResNetD captures the nonlinear relationship between input measurements and output states. The output states produced by base-learner ResNetD are very close to actual states and they capture various features existing between input measurements and output states. The multivariate linear regression (MLR) is used to form the ensemble model for estimating the final power system states (voltage magnitudes and phase angles). The trained ensemble learning  model is used to predict power system states in real-time. During testing phase, additional Gaussian noise is added in the data to test the robustness of the proposed approach against measurement errors. During the implementation of the proposed data-driven PSSE in real-time, there may be missing measurements that lead to the failure of the state estimation. To deal with this problem, we adopted the MLR to forecast missing states at any instant. The accuracy and efficiency of the proposed method against standard ML methods is validated through comprehensive case studies on the IEEE $14$, $30$, $57$, $69$, and $118$ bus benchmark systems.


The major contributions of this paper toward ML-based state-of-the-art state estimation are summarised as follows.
\begin{itemize}
    \item 
    Deep neural networks have the capability to map nonlinear relationships between the input data and the output because of their nonlinear nature. They provide great flexibility and scalability with the system size and amount of available samples. However, deep neural networks learn through stochastic training algorithms, which results in high variation in training parameters of the model. This may make deep neural networks find different sets of weights every time they are trained and may produce different results. This work proposes ensemble learning setup to solve the high variance problem associated with the state-of-the-art deep learning based state estimation techniques. Ensemble learning models train multiple models and combine the output of those models for the final prediction which results in variance reduction. 
    The ensemble learning model not only reduces the variance in prediction but also its performance improves in terms of accuracy and efficiency if the models are selected appropriately.

    \item Motivated by the capability of recently advanced residual neural net-work architectures \cite{8099726, 8372953, 8754766, 10.5555/3304889.3305101, HANIF202028, ZOU201939} to map nonlinear relationships between the input and the output variables, the ResNetD is developed to capture the nonlinear relationship existing in the state estimation problem.  Also, the work presented in \cite{4074246, 6345595, 7312519, 4643662} for state forecasting shows that linear models can appropriately forecast power system states using historical states.  Therefore, this paper utilises a number of ResNetD models as base-learners to predict states that act  similar  to  historical  states  used  for  state  forecasting  approaches. However, states predicted by ResNetD are much closer to actual states due to its capability to map the nonlinear relationship between input measurements and system states. MLR maps the relationship between the outputs of the base-learner modes and the actual states to further improve the overall performance.
\end{itemize}

The rest of the paper is organised as follows. Section \ref{pformulation} presents details on power system state estimation and problem formulations. Section \ref{ensemble_psse_estimator} describes the proposed deep ensemble learning setup for PSSE with an algorithm to deal with missing measurements.  Section \ref{simulation} examines the proposed approach through numerical case studies. Section \ref{conclusion} provides concluding remarks.

\section{PSSE Problem Formulation}\label{pformulation}
This section briefly discusses the preliminaries and problem formulation of PSSE. We will not reproduce rigorous derivations of the PSSE problem; rather, we will use the expressions of the PSSE problem to develop the proposed approach.
\subsection{Preliminaries of PSSE}
Given network configuration and parameters and a set of measurements, $z$, the AC state estimation determines system states as follows \cite{PSSEALI2004}. 
\begin{equation}\label{equ:state_estimation}
    z=h\left(x\right)+e,
\end{equation}
where \\
 $z=[z_1, z_2, \cdots, z_m]$: set of measurements\\
 $x=[x_1, x_2, \cdots, x_{2n}]$: vector of state variables\\
 $e=[e_1, e_2, e_3, \cdots, e_m]$: vector of measurement residuals\\
 $h=[h_1(x), h_2(x), \cdots, h_m(x)]$: nonlinear function (i.e., system model) relates state vector to the measurement set.

Historical real power system measurements and states are not easily accessible for training and testing of the proposed data-driven PSSE. Therefore, WLS with Gauss-Newton method is used to generate training data. WLS-based optimisation to determine the estimated state vector, $\hat{x}$, can be expressed as follows.
\begin{equation}\label{equ:state_objective}
    \mbox{min } J(x)=\frac{1}{2}\left(z-h\left(x\right)\right)^T W\left(z-h\left(x\right)\right),
\end{equation}
where $W$ is the weight vector developed based on the variance of the measurement errors ($\sigma_1^2, \sigma_2^2, \cdots, \sigma_m^2$) represented as,
\begin{equation}   W=\begin{bmatrix}\textstyle \frac {1}{\sigma _{1}^{2}} & 0 & \cdots & 0\\\textstyle 0 & \frac {1}{\sigma _{2}^{2}} & \cdots & 0\\\textstyle \vdots & \vdots & \ddots & 0\\\textstyle 0 & 0 & \cdots & \frac {1}{\sigma _{m}^{2}} \end{bmatrix}\! \end{equation}

The minimum value of the optimisation problem \eqref{equ:state_objective} can be computed using a first-order optimality condition as follows. 
\begin{equation}\label{eq:first_order_derivative}
    g(x)=\frac{\partial J(x)}{\partial x}=-H^T(x)W\left[z-h(x)\right]=0\mbox{.}
\end{equation}

The state vector $\hat{x}$ in \eqref{eq:first_order_derivative} can be solved as the limit of the sequence of states, $\hat{x}_k$, by means of Gauss-Newton recursive scheme; one step of such recursive scheme can be presented as follows.
\begin{equation}
    \hat{x}_{k+1}=\hat{x}_k+G(\hat{x}_k)^{-1}H^T(\hat{x}_k)W[z-h(\hat{x}_k)]\mbox{,}
\end{equation}
where $H$ and $G$ are, respectively, Jacobian and Gain matrices and can be expressed as follows.
\begin{gather}
    H(\hat{x_k})=\Big[\frac{\partial h(x)}{\partial x}\Big]_{x=\hat{x}_k}\mbox{,}\\
    G(\hat{x}_k)=H^T(\hat{x}_k)WH(\hat{x}_k)\mbox{,}
\end{gather}
where $\hat{x}=[\hat{V_t}^1, \hat{V_t}^2,....,\hat{V_t}^n, \hat{\delta_t}^1, \hat{\delta_t}^2,....,\hat{\delta_t}^n]$ is the estimated state vector---voltage magnitudes, $\hat{V_t}^i$, and phase angles, $\hat{\delta_t}^i$, for the $i$\textsuperscript{th} bus at time $t$.

\subsection{Problem Statement}
The problem of data-driven PSSE is to map the available set of measurements, $z_t$, to the power system state variables, $\hat{x}_t$. This problem can be expressed as follows. 
\begin{equation}\label{eq:non_lin_rel_xz}
    \hat{x}_t=f(z_t)\mbox{,}
\end{equation}
where the set of measurements can be expressed as follows.
\begin{gather*}
\begin{aligned}
z_t=&\left[\{|V_t^i|\}_{i\in \zeta_t^v}, \{\delta_t^i\}_{i\in \zeta_t^{\delta}} \{P_t^i\}_{i\in \zeta_t^p},\right.\\ 
&\left. \{Q_t^i\}_{i\in \zeta_t^q}, \{P_t^{ii'}\}_{(i,i)\in \pounds_t^p },  \{Q_t^{ii'}\}_{(i,i)\in \pounds_t^q }  \right]^T
\end{aligned}
\end{gather*}
with $\{|V_t^i|\}$ being voltage magnitude measurements at any instant $t$ that are available in $\zeta_t^v$ buses;  $\{\delta_t^i\}$ is phase angle measurements at any instant $t$ that are available in $\zeta_t^{\delta}$ buses; $\{P_t^i\}$ and $\{Q_t^i\}$ are real and reactive power injection measurements at any instant $t$ that are available in  $\zeta_t^p$ and $\zeta_t^q$ buses, respectively; and $\{P_t^{ii'}\}$ and $\{Q_t^{ii'}\}$  respectively, are power flow measurements from bus $i$ to $i^{'}$ at any instant $t$ that are available in $\pounds_t^p$ and $\pounds_t^q$ lines. ($\zeta_t^v$, $\zeta_t^{\delta}$, $\zeta_t^p$, $\zeta_t^q$)$\in \zeta$ with $\zeta=\{1, 2, \cdots, n\}$ being the set of all buses and ($\pounds_t^p$, $\pounds_t^q$) $\in \pounds$ with $\pounds$ is total number of branches in a power network. 

The function, $f$, contains weights ($w_1$, $w_2$, $w_3$, $\cdots$, $w_2n$) that map the relationship between input measurements and output states. The problem is to find the weights, $w_1, w_2, w_3, \cdots, w_{2n}$, that reduce the overall loss between predicted and actual states. 

\section{The Proposed PSSE Method}\label{ensemble_psse_estimator}
In the proposed method, an neural network in the ensemble learning setup is used. Before proceeding further to describe the proposed model, the functionality and importance of the ensemble learning in solving the PSSE problem are explained as follows. 

In ensemble learning, multiple machine learning algorithms are brought out together to solve the same prediction or classification problem. Subsequently, results from different methods are collected and combined. Machine learning (ML) models in the ensemble learning are called base-learners that weakly predict a certain parameter. These weak learners are trained to generate a set of hypotheses and subsequently combined to produce more accurate results. The base-learners are combined either in a sequential or parallel manner. The final results are obtained using various techniques including, but not limited to, majority voting, averaging, and weighted averaging \cite{SPRINGERZhang2012}.

Traditional ML models suffer from two disadvantages. First, given a training data set, it is often not possible to find the best ML algorithm due to their black-box nature. Thus, although data-driven ML models provide superior results over other models, this performance cannot be explained \cite{BARREDOARRIETA202082}. In other words, regardless of how many times these models are tried (e.g., in a trial and error fashion), users may not be able to identify the best model. Second, even if the best algorithm is identified, the ML model may not provide the optimal performance for certain sample data as the search process of disparate ML algorithms is different. Thus, to compensate for the error of some models, it is feasible to combine different learners to get optimal performance.  The capability of ensemble learning lies in the fact that the base learners are diverse in nature. The diversity can be obtained using different ML models, different training parameters, different training data sets, and of course, a combination of all of them \cite{SPRINGERZhang2012,Lappalainen2000}. The added diversity enables the models to correct errors of some members as different learners make different errors on the same set of inputs. However, careful attention is needed while forming the variations in  models, datasets, or training parameters. Selection of inappropriate base-learner models may worsen  the  performance if the majority of  selected base-learners capture similar features and miss a critical feature even if a few other base-model capture the critical features. Thus, the combination of carefully chosen different base-learners reduces the overall prediction error. Hence, ensemble learning is a powerful method to ensure accurate generalisation capability in a training process.

The major drawback of the ensemble learning is that, as compared to a single base learner, the model is redundant and requires more training. However, neural network training is a one-time offline process. Thus, ensemble learners provide a generalised model, but it requires extra training (that is done a single time). Since the training process is done only once, it does not pose a problem when a system operates. In the operating phase, the execution time is still low (operating phase execution time for the proposed model is provided in result section) and better than the mathematical model based state estimators.

\subsection{Attributions of the Proposed Model}\label{base_learners}
This paper has used a stacking ensemble technique to develop ensemble learning model for PSSE. In the stacking ensemble technique, the base-learners are combined in  parallel.  In this technique, heterogeneous weak learners learn on the training data independently. The independent learners are combined using a meta-model that provides output based on the predictions obtained from weak learners  \cite{SPRINGERZhang2012}. This paper uses dense residual neural network (ResNetD) model as a base learner. Before explaining the functionality of ResNetD, we describe the development of the proposed ResNetD as a base-learner.

Many architectures of the neural network have been proposed in the literature to map the non-linear relationship between input and output vectors of a given system. This includes, but not limited to, classical ML techniques (e.g., decision tree and k-nearest neighbours regression), multilayer perceptron (MLP), CNN, recurrent neural network (e.g., LSTM), and hybrid architectures (e.g., CNN-LSTM and ResNet). Given the aforementioned advantage of ensemble learning, we leverage ensemble learning setup for PSSE. To develop an appropriate model of ensemble learning setup for PSSE, ML models including MLP, CNN, LSTM, and ResNet are tested as base learners. A number of these models individually and in combination are stacked in parallel to test their performance with various meta-learners. ML models, such as MLP, ResNet, k-nearest neighbours, decision tree, CNN, and MLR, are tested as meta-learner for the above-mentioned base-learners. Individually, dense neural network-based ResNet (ResNetD) architecture appropriately maps the non-linear relationship between input measurements and the output state variables than any other approach for our problem. Development and testing of ResNetD is motivated from recently advanced residual neural network architectures \cite{8099726, 8372953, 8754766, 10.5555/3304889.3305101, HANIF202028, ZOU201939} to  map nonlinear relationship between the input and output variables. Also, with above-mentioned models as meta-learner, ResNetD has produced better result compared to other architectures. The results produced by the combined architecture of ResNetD as base-learners are very close to true states. Authors of \cite{4074246, 6345595, 7312519, 4643662} have demonstrated that linear models can accurately forecast power system states using historical states. Since results obtained from ResNetD as base-learners are similar to historical states (but much close to actual states than historical states because of the capability of ResNetD to capture the nonlinear relationship) used for state forecasting approaches, MLR maps these states even closer to actual states. Therefore, a number of parallel ResNetD as base-learners and MLR as meta-learner is taken as ensemble learning setup for the PSSE problem.

\subsection{Residual Neural Network as a Base Learner Model}\label{base_learners}
Residual Neural Network (ResNet) is a type of artificial neural network that builds on a structure known from pyramidal cells in the cerebral cortex. ResNet is formed by skipping the connections or by jumping over some layers of the feed forward neural network. Typical ResNet is formed by skipping two or three layers that contain batch normalisation and a non-linear function (rectified linear unit (ReLU)) in between. These skipped connections are important in ``vanishing'' and ``exploding'' gradient issues by reusing activation function from a previous layer until the adjacent layer learns its weights \cite{7780459, 8099726}. Another advantage of skipping layers is that it simplifies the network and speeds up learning processes as fewer layers are used in the training. 

\begin{figure*}
    \centering
    \includegraphics[scale=0.345]{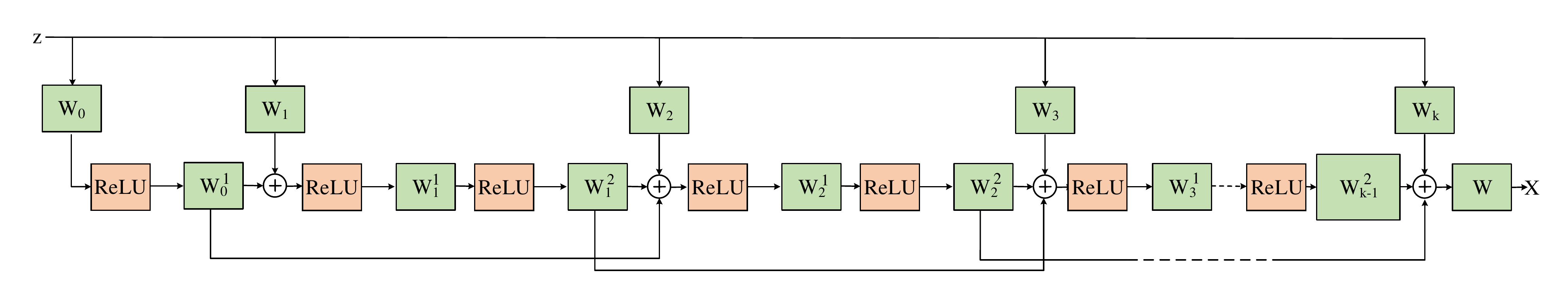}
    \caption{Architecture of ResNetD with $K=2$, where $K$ denotes number of hidden units $W$, and $Z$ and $X$ represent the input measurement vector and output state variable vector, respectively.}
    \vspace{-2ex}
    \label{fig:base_ResNet}
\end{figure*}

Inspired by ResNet architecture proposed in \cite{8099726, 8372953, 8754766, 10.5555/3304889.3305101, HANIF202028, ZOU201939}, a ResNetD is developed as shown in Fig.~\ref{fig:base_ResNet} as a base-learner. One block of the proposed ResNetD architecture is formed by merging the regular information flow, the output of previous blocks' dense layers, and connecting the input through a dense layer directly (as shown in Fig.~\ref{fig:base_ResNet} with $2$ skip neurons in the regular information flow). The advantage of this approach is that it improves the information flow and recovers the missing features. In this paper, Hubber loss is employed as loss function because of its robustness against outliers \cite{Blmann2014}. ReLU is used as an activation function for the proposed ResNetD. 


\subsection{The Proposed Deep Ensemble Learning Setup}\label{ensemble} 
\begin{figure*}[ht!]
\centering
    \includegraphics[scale=0.8]{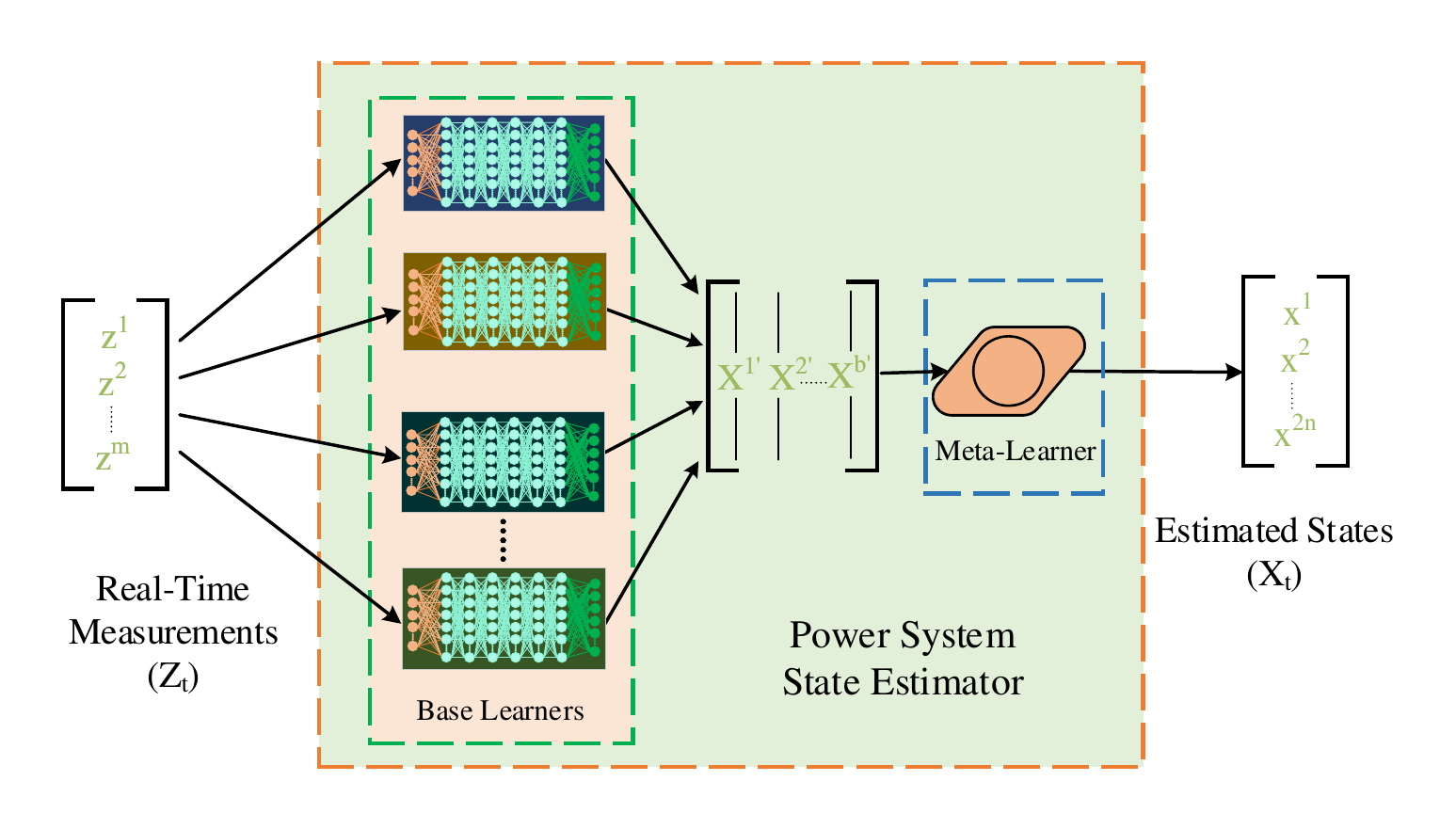}
    \caption{Proposed deep ensemble learning based PSSE.  Input vector $Z_t$ is the real time measurement vector of size $m$. The intermediate vectors $X^{1'}, X^{2'}\cdots X^{b'}$ are the estimated voltage magnitudes and phase angles from $b$ number of base-learners. The output vector $X_t$ includes $n$ voltage magnitudes and $n$ phase angles estimate for a power network with $n$ buses.}
    \label{fig:ebnsemble_dnn}
\end{figure*}

The proposed model uses stack generalisation of the ensemble learning to predict power system states. The proposed architecture employs a number of parallel ResNetD as base-learner and MLR as meta-learner. Although all of the six ResNetD models used in this paper have the same architecture, they act like a diverse set of models because of the stochasticity involved in the model. Therefore, even though the models are redundant, their outputs will be different and the differences in their outputs result in the formation of appropriate base-learners. The architecture of a base-learner, ResNetD, is provided in Section \ref{base_learners}. Brief description of MLR is provided in \ref{MLR}, as it is well-known technique. Fig.~\ref{fig:ebnsemble_dnn} shows the basic architecture of the proposed PSSE. 

It is assumed in the proposed work that similar operating patterns to current state exist in the historical dataset. Similar operating patterns do not always mean the same topology. For large power systems, change of topology at the local level may not change the electricity generation or consumption in the local network. Operating points may still be considered similar to historical points if the local network is considered as an aggregated node \cite{7378513}. For the case of topology change at a higher level in the bulk power system, where the operating point may change with the change of topology, the proposed model has to be trained again with the historical dataset of the changed topology. For the bulk power system, the immediate training after the topology change may be computationally challenging and expensive. The frameworks presented in \cite{nagaraj2020ensemble, 9176958} work with adaptive learning and could be used for the case of topology change at a higher level in the bulk power system. However, further detailed analysis is needed to deem the applicability of approaches presented in \cite{nagaraj2020ensemble, 9176958} in our proposed work, which is left as a future work. These assumptions also exist in most of the ML based state-of-the-art state estimation approaches. Although it has been shown in \cite{5233865, JI2021106412, 8091099} that the forecasting aided state estimator can address the lack of measurements,  measurement errors, grid topology and link parameters change, the problem of topology change in the bulk system that changes the operating point still a challenge.

For  normal  conditions, the  proposed state estimation model can accurately predict power system states. If some of the measurements are missing or time delayed, the forecasted states obtained from the proposed state forecasting approach can be utilised to estimate the missing measurements as pseudo-measurements. Forecasting-aided state estimation approach can also deal with measurement errors, network configuration, sudden changes in the network, and change in network parameters \cite{5233865}. For a state estimation approach to be robust, it must be insensitive to major measurements errors and network topology changes (\cite{PSSEALI2004} chapter 6). Therefore, the proposed state estimator is robust against local topology change, the missing measurements, and the measurement errors.

The purpose of the PSSE is to estimate voltage magnitudes and phase angles at all $n$ buses of a power system at any instant using measurements obtained from various measurement devices. In practical power systems, measurements can come from different measurement devices including PMU and SCADA. Also, different measurement types and locations introduce time synchronisation and time skewness issues because of the different latency of the measurements. Several approaches, for example \cite{8494764, KORRES20111514, 6338328, 8248787}, have been presented to deal with this challenge. In this paper, we have assumed that measurements are synchronised using one of already available synchronisation approaches. The measurements, $z^T=[z_1, z_2, \cdots, z_m]$, obtained from the field devices may consist of real and reactive power flows in different branches, nodal voltage magnitudes and phase angles, and real and reactive power injections at various buses of a power network.

\begin{figure}
    \centering
    \vspace{-4ex}
    \includegraphics[scale=0.45]{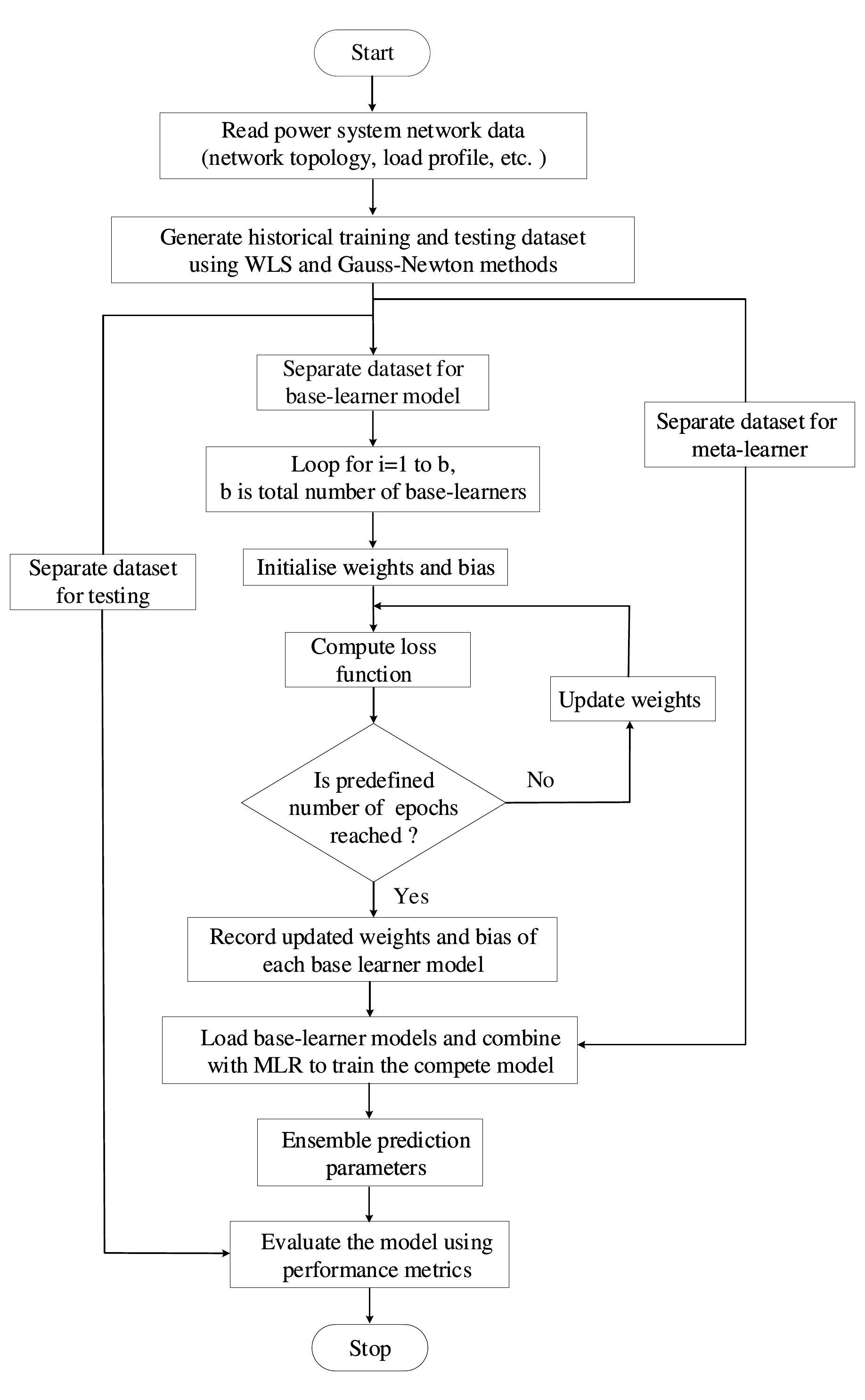}
    \caption{Flowchart of the proposed PSSE. Huber loss is the used as loss function. The proposed work is run for $200$ epochs, this number is determined empirically by looking into the training of ML model error settling around.}
    \label{fig:flowchart}
\end{figure}

It can be seen from Fig.~\ref{fig:ebnsemble_dnn}, in the proposed approach, $m$ measurements (with $m\geq 2n$ as the necessary condition for the system to be observable; observability of a network depends upon several conditions including type and location of measurements as well as the network topology, the details on the system observability condition are provided in chapter 4 of \cite{PSSEALI2004}) are used as inputs to the base-learners. Each of the base-learners computes the state vector independently (parallel stacking) as an output vector. The output of the base-learners is provided as input to the meta-learner that predicts the final state vector variable, $X$, which consists of $n$ voltage magnitudes and $n$ phase angles for $n$ buses of a power network. The meta-learner (MLR) maps prediction very close to actual states. Hereinafter ``Stacked ResNetD'' is used to denote this ensemble learning model. Fig.~ \ref{fig:flowchart} shows the flowchart of the proposed method.

For the training and testing of the proposed model, the availability of a complete set of historical measurements and states is assumed. While at the instant of real-time operation of the proposed PSSE, if some of the measurements are missing or topology are changing, the forecasted states can be used for the monitoring and control of the power system. Forecasting aided state estimation helps to deal with errors, sudden change in the network, and topology and network parameters change \cite{5233865,8091099, 8754766, 8683139, JI2021106412}; therefore, forecasting missing measurements is important for the completeness of the data-driven PSSE. Multivariate-linear regression for state forecasting is described in section \ref{MLR} to deal with the problem of missing measurements.

\subsection{Multivariate-linear Regression for State Forecasting}\label{MLR}
The general assumption of data-driven PSSE approaches is the availability of complete measurements during real-time state estimation, which may not be true in real-time data acquisition. Some of measurements may not be available while performing state estimation at control centres due to various reasons such as measurement device failures, unreported outages, denial of the service attacks, transmission line sags, and transmission channel failures. Although missing measurement data may not be frequent in practical power systems, an algorithm is needed for missing measurements for the completeness of the proposed state estimation approach. To deal with these missing measurements/states, we forecast missing states. Forecasting of missing states has several advantages because it compensates for various errors, sudden changes, and topology and network parameter changes. The advantages of forecasting missing states or measurements to deal with changes in network typologies and  parameters are provided in \cite{5233865}. Mathematically, forecasting system states for next hours ($x_{t+1}$) given historical states ($x_{t-h}, x_{t-h+1}, ......., x_{t-1}, x_t$) can be expressed as follows. 
\begin{gather}
    \hat{x}_{t+1}=\phi (x_{t-h}, x_{t-h+1}, ......., x_{t-1}, x_t) \label{equ_vf}\\
    \hat{z}_{t+1}=h_{t+1}(\hat{x}_{t+1})+e_{t+1} \label{z_hatprediction}
\end{gather}
It can be seen from \eqref{equ_vf} that an appropriate mapping function $\phi$ needs to be developed to appropriately forecast next state. Forecasted states can be used directly as states of next hours or can be used in \eqref{z_hatprediction} as states to generate the pseudo-measurements of the next hour. The missing measurements are replaced with the pseudo-measurements and provided as input to the proposed state estimator along with the available measurements. 
The performance of the linear and non-linear models are similar for the state forecasting \cite{4074246, 6345595, 7312519, 4643662}. However, nonlinear models come with added computational time and complexity. Therefore, MLR is adopted in this work to forecast future states. MLR has following benefits over the other non-linear models: it is easy to train; takes much less time for training; and is much easier to understand. The input to MLR is historical states as explained above and the output is the states of the next one step (one hour if the state estimation is done hourly) or  more hours (for multi step forecast). In this work, most recent $24$ hours of the historical time-series states are used as input and output states of the next hour is forecasted. Analytically, MLR can be expressed as follows. 
\begin{gather}
    \hat{x}_{t+1}=\alpha_{0} + \beta_{0}x_t + \beta_{1}x_{t-1} + ..... + \beta_{h-1}x_{t-h+1} + \beta_{h}x_{t-h} \label{mlr_v}
\end{gather}
where $\alpha_0$ is the intercept and $\beta_{0}$ through $\beta_{h}$ are regression coefficients for state forecasting.

\subsection{Evaluation Metrics}\label{evaluation_metrics}
The proposed work is compared with various state-of-the-art methods using the following evaluation metrics.
\begin{enumerate}
    \item Mean Absolute Error (MAE):
    \begin{equation}\label{}
        MAE=\frac{1}{TN}\sum_{t=1}^T\sum_{i=1}^{N}|x_t^i-\hat{x}_t^i|
    \end{equation}
    \item Root Mean Square Error (RMSE):
    \begin{equation}\label{}
        RMSE=\sqrt{\frac{1}{TN}\sum_{t=1}^T\sum_{i=1}^{N}(x_t^i-\hat{x}_t^i)^2}
    \end{equation}
\end{enumerate}
where $N=n$ is the total number of estimated voltage or phase angle states; $T$ is the total number of test samples; and  $x_t^i$ and $\hat{x}_t^i$ represent actual and predicted states, respectively.

\section{Numerical Evaluation}\label{simulation}
This section validates the proposed approach through numerical evaluation. Section \ref{data_preparation} describes step-by-step training and testing data generation. Section \ref{perf_stacked_resnetd_psse} presents comparisons between the proposed Stacked ResNetD structure and existing methods including MLP, CNN, and ProxlNet for PSSE. Section \ref{sec:non_gaussian_noise} shows the performance of the proposed model with Non-Gaussian Noise in the measurements. Section \ref{GM_compare} shows the capability of proposed PSSE to emulate  generalized maximum-likelihood (GM)-estimator proposed in \cite{7742899, 496203} for power system state estimation. Finally, section \ref{perf_MLR_forecast} shows the performance of the MLR against CNN, LSTM, and CNN-LSTM for time-series state forecast. 

\subsection{Dataset Generation}\label{data_preparation}
The performance of the proposed  state estimation is demonstrated through various case studies on the standard IEEE benchmark systems: IEEE $14$, IEEE $30$, IEEE $57$, IEEE $69$, and IEEE $118$-bus systems. As historical measurements and states of real power systems are not easily accessible for training the proposed method, training data are generated using real power load dataset (varying power demands help capture the dynamics of real power systems) from Global Energy Forecasting Competition $2012$ \cite{HONG2014357}. Load profile of zone $1$ is taken in this paper. Load profiles are normalised to match the scale of the tested systems. The load time series dataset is normalised by the peak value as follows. Let $X={x_1, x_2, \cdots, x_{8760}} $ denote the actual load in the dataset. The normalised load profile can be determined as follows. 
\begin{equation}\label{normalized}
X_{profile}=\frac{x_1, x_2, ......., x_{8760}}{\text{max}(x_1, x_2, ......., x_{8760})}
\end{equation}
To obtain the load demand at each node for each time instant, actual load demand at each node is multiplied by the normalised load profile of \eqref{normalized}.

Simulations are performed using Power System Simulation Package (MATPOWER) \cite{5491276} to generate the data and Python with Keras and Scikit-learn libraries is used for training and testing of machine learning models.  AC power flow is run for the entire simulation period of the load data and various power flow results such as line flows (real and reactive power flows), nodal voltage magnitudes and phase angles, and nodal power injections are recorded. Gaussian and non-Gaussian noises are used to emulate real-world data. While estimating system states from measurements using WLS method to generate the training and testing data, measurement standard errors ($0.01$, $0.01414$, $0.01414$, and $0.0122$) are used for voltage magnitudes, phase angles, line power flows, and nodal power injections, respectively. One of the necessary conditions for a system to be observable is that the total number of measurements should be greater than or equal to the total number of states to be estimated.

The number of measurements taken for the 14-bus system is  $64$. Although it can be predicted only with $32$ measurements (data for this case are provided in the shared code), 64 measurements are provided because having redundant measurements have several benefits such as (a) it improves the performance of the model when there are suspected measurements; (b) it can obtain better estimate for the suspected data sets;  (c) has capability to estimate important non-telemetered variables (e.g., transformer taps); and capability to determine the unknown status of CBs and to detect topological errors \cite{PSSEALI2004}. For the $30$-bus system, $110$ measurement data points are used; the proposed methods can also with work with less number measurements, for example $80$ measurements (data for this case are provided in the shared code). For $57$-, $69$-, and $118$-bus systems, $216$, $210$, $562$  number of measurements are chosen, respectively. For this study, the number of measurements are determined empirically.

As an example of location of measurement devices, the locations of the $32$ measurements for the $14$-node system are as follows: bus real and reactive power injection measurements are taken from buses $[$2, 4, 8, 10, 11, 12, 14$]$, i.e., total of $7\times 2=14$ measurements; voltage magnitude and phase angle measurements of bus $1$; i.e., total of $1\times 2=2$ measurements; and real and reactive power flow measurements are taken as follows (from bus--to bus): $1$--$2$ ,$2$--$3$, $2$--$5$, $4$--$6$, $4$--$7$, $6$--$11$, $6$--$13$, $12$--$13$, i.e., $2\times 8=16$. Locations for the tested systems are determined empirically and by following a similar location as that of state estimation work presented in \cite{7742899, 496203}. 

Measurement locations of the SCADA system is very important as the number of measurements and locations have influence over the result and observability of the system. However, determining the optimal number of measurements and optimal locations of SCADA is outside the scope of this paper.

\subsection{Results of Stacked ResNetD for PSSE}\label{perf_stacked_resnetd_psse} 
The $m$ measurements are provided as input and the $n$ voltage magnitudes and $n$ phase angles are provided as estimated outputs to train the model which can be performed off-line for a real power system. The trained model can be used in real-time to estimate current states of the system with given current measurements. 

For each of the IEEE $14$, $30$, $57$, and $69$ bus test systems, a total of $39,444$ data points are generated. For the IEEE $118$ bus system, a total of $17,520$ data points are generated. From the total dataset,  $40\%$ are utilised for training the base-learners, $36 \%$  for training the meta-learner, and the remaining $24\%$ are used for testing the complete ensemble learning setup. The point to be noted while training the meta-learner is that it must be trained with separate data-set than the one used to train base-learners to avoid the over-fitting. A Gaussian white noise with signal-to-noise ratio of $50$ dB is added to the training data set. In real-time measurements, errors at any instant may be different from that of the previous instant of time. To capture changing measurement errors and check the robustness of the proposed model against measurement errors, Gaussian white noises with signal-to-noise ratio of $20$ dB are added in the test data set to alter them more than training data set. Gaussian noise is considered based on the general convention used to generate a dataset for data-driven based state estimation. However, real measurements do not necessarily follow the Gaussian noise. Specifically, load does not usually follow a Gaussian distribution. Advance metering infrastructure can be used to develop distribution functions for load points. To test the proposed approach on different distributions and noises, we have used non-Gaussian noise as well in section \ref{sec:non_gaussian_noise}. The per-unit values of voltage magnitudes are converted to percent values and the phase angles are converted from radian to degree for better visualisation. 

The proposed Stacked ResNetD for state estimation is compared with multi-layer-perceptron (MLP), CNN, and Prox-linear net.
\begin{itemize}
    \item Multilayer Perceptron (MLP): MLP has a layered architecture with input, hidden, and output layers. The normalised input is fed at the input layer. The cardinality of the input vector determines the number of neurons in the input layer. There can be multiple hidden layers in the MLP. The final prediction output is obtained from the output layer. In this work, MLP consisting of $6$ hidden layers with ReLU as activation function and adaptive moment estimation (Adam) as optimisation function is used.
    \item  Convolution Neural Networks (CNNs): CNNs are often used in many applications. The CNNs have a convolution layer followed by a pooling layer. The CNN and pooling layers find the low level feature of the input vector. Fully connected layers are added after the convolution and pooling layers. The CNN architecture is well-suited for 2-D input. However, it can also be used efficiently for 1-D inputs like time-series. One of the benefits of CNN is that they are easier to train and have a fewer parameters as compared to the fully connected neural network with the same number of hidden units. The CNN architecture is presented to compare it with the proposed approach which consists sequentially of: two 1-D convolution layer with $64$ filters and kernel size of $3$; one 1-D max pooling layer with pool size of $3$; one 1-D convolution layer with filter size of $128$ and kernel size of $3$; one 1-D global average pooling layer; two dense layers with $4n$ units and ReLU activation function; and a output dense layer with $2n$ units and ReLU activation function. CNN model uses Adam as an optimiser.
    \item Prox-linear Net (ProxlNet): Authors of \cite{8754766} have proposed a ProxlNet for real-time PSSE. The ProxlNet has been formed by skipping the layer that connects the input directly to the immediate output layer, where each layer consists of a fixed number of hidden layers. The ProxlNet architecture used for comparison consists of $2$ skip-connection layers with $3$ hidden units between each layer. For the detailed architecture of ProxlNet, refer to \cite{8754766}. 
\end{itemize}

The aforementioned existing techniques are run $6$ times independently with a batch size of $64$ and $200$ epochs, and the minimum values (prediction vary because of the stochastic nature of the deep learning models) of RMSE and MAE  for all runs are taken for comparison with the proposed Stacked ResNetD. The batch size of $64$ and $200$ epochs are determined empirically. The number of epochs are selected after observing the training error settling in the machine learning models. The number of epochs could be different for different models; however, in this work, we have determined it conservatively. In other words, during the training phase, some of models may settle earlier than 200 epochs while others may take around 200 epochs to settle; therefore, the number of epochs is chosen to be $200$. As the training is offline procedure, the number of training epochs can be selected based on system requirements. The structure of ResNetD used as a base-learner is same as shown in Fig.~\ref{fig:base_ResNet} where three blocks containing $2$ skip hidden units in the regular information flow of each blocks are used. The number of neurons selected for each input and hidden layers is the total number of input measurements of a specific system; for the output layer, the total neurons equal to total number of states to be predicted.

As the size of the data is big, we  train all $6$ parallel ResNetD base-learners with same dataset and parameters. Before deciding to train all the base-learners with same set of training, we also tested the performance of ResNetD as base-learners by dividing the training data into $6$ folds. The performance of Stacked ResNetD on the test dataset is better when trained with the entire training dataset for all of the parallel base-learners than that when trained with $6$ fold  of data for $6$ parallel ResNetD.
\begin{table*}[h!]
\caption{Comparison of MLP, CNN, ProxlNet, and proposed Stacked ResNetD in terms of RMSE and MAE metrics for voltage magnitudes estimation}
\centering
\begin{tabular}{|l|l|l|l|l|l|l|l|l|l|l|}
\hline
\multirow{2}{*}{Models} & \multicolumn{2}{l|}{IEEE $14$ Bus} & \multicolumn{2}{l|}{IEEE $30$ Bus} & \multicolumn{2}{l|}{IEEE $57$ Bus} & \multicolumn{2}{l|}{IEEE $69$ Bus} & \multicolumn{2}{l|}{IEEE $118$ Bus} \\ \cline{2-11} 
& RMSE          &     MAE      & RMSE           &  MAE          & RMSE           &    MAE & RMSE           &  MAE          & RMSE           &    MAE      \\ \hline \hline
MLP                  &     $2.4574$  &    $1.8533$    &  $4.1737$         & $2.9253$          & $5.1976$          & $3.5885$ & $6.8272$ & $5.3114$& $1.8588$& $1.3944$          \\ \hline

CNN                  &     $2.2262$  &    $1.4926$    &  $4.2576$         & $2.9005$          & $5.1595$          & $3.4571$ & $6.8614$ & $5.3225$& $1.9343$& $1.4635$          \\ \hline
ProxlNet                  &     $2.4592$  &    $1.8815$    &  $4.1405$         & $2.8885$          & $5.1273$          & $3.5283$ & $6.6526$ & $5.1737$& $1.8385$& $1.3784$          \\ \hline
Proposed
 &       $\mathbf{0.2605}$    & $\mathbf{0.1660}$          &  $\mathbf{0.4753}$         & $\mathbf{0.2852}$          & $\mathbf{0.4766}$          & $\mathbf{0.2931}$ & $\mathbf{0.6486}$ & $\mathbf{0.4238}$& $\mathbf{0.1894}$& $\mathbf{0.1196}$          \\
  Stacked ResNetD &&&&&&&&&& \\
 \hline
\end{tabular}
\label{tab:comparisonDvoltage}
\end{table*}

\begin{table*}[h!]
\caption{Comparison of MLP, CNN, ProxlNet, and proposed Stacked ResNetD models in terms of RMSE and MAE metrics for phase angles estimation}
\centering
\begin{tabular}{|l|l|l|l|l|l|l|l|l|l|l|}
\hline
\multirow{2}{*}{Models} & \multicolumn{2}{l|}{IEEE $14$ Bus} & \multicolumn{2}{l|}{IEEE $30$ Bus} & \multicolumn{2}{l|}{IEEE $57$ Bus} & \multicolumn{2}{l|}{IEEE $69$ Bus} & \multicolumn{2}{l|}{IEEE $118$ Bus} \\ \cline{2-11} 
& RMSE          &     MAE      & RMSE           &  MAE          & RMSE           &    MAE & RMSE           &  MAE          & RMSE           &    MAE      \\ \hline \hline
MLP                  &     $0.5496$  &    $0.3533$    &  $1.2202$         & $0.6854$          & $1.8197$          & $1.1970$ & $2.5232$ & $1.6724$& $1.4351$& $1.0852$          \\ \hline
CNN                  &     $0.6496$  &    $0.4631$    &  $1.2513$         & $0.7312$          & $1.8677$          & $1.2741$ & $2.5242$ & $1.6980$& $2.1926$& $1.5885$          \\ \hline
ProxlNet                  &     $0.5122$  &    $0.3213$    &  $1.1840$         & $0.6581$          & $1.7650$          & $1.1558$ & $2.4475$ & $1.6324$& $1.3061$& $1.0045$          \\ \hline
Proposed                   &       $\mathbf{0.1102}$    & $\mathbf{0.0733}$          &  $\mathbf{0.5472}$         & $\mathbf{0.1873}$          & $\mathbf{0.2978}$          & $\mathbf{0.1581}$ & $\mathbf{0.5417}$ & $\mathbf{0.3156}$& $\mathbf{0.2104}$& $\mathbf{0.1272}$          \\ Stacked ResNetD &&&&&&&&&& \\ \hline
\end{tabular}
\label{tab:comparisonDphase}
\end{table*}

Table \ref{tab:comparisonDvoltage} shows comparison of MLP, CNN, ProxlNet, and proposed Stacked ResNetD models in terms of RMSE and MAE of voltage magnitude estimation for IEEE $14$, $30$, $57$, $69$, and $118$ benchmark systems. The values of the metrics show that the proposed Stacked ResNetD ensemble learning setup captures the true relationship between input measurements and the estimated voltage states. The proposed base-learner has regular information flow, skipping connection, and direct connection to the input data through dense layer.  All this together solving ``vanishing" and ``exploding" issues, improving the regular information flow and recovering missing features. MLR as meta-learner is further improving the result toward the actual values.

Table \ref{tab:comparisonDphase} shows the comparison of MLP, CNN, ProxlNet, and the proposed Stacked ResNetD in terms of RMSE and MAE metrics of phase angles estimation for IEEE $14$, $30$, $57$, $69$, and $118$ benchmark systems. The results obtained by the proposed approach is closer to actual state values.

The run-time performance of each model is determined over all the testing dataset and is averaged over each instance. Table~\ref{tab:Runtime} shows the run-time performance of each model per instance estimation. 
\begin{table*}[h!]
\caption{Comparison of MLP, CNN, ProxlNet, and proposed Stacked ResNetD models in terms of run-time performance per instance estimation where `s' denotes seconds and `ms' denotes milliseconds.}
\centering
\begin{tabular}{|l|l|l|l|l|l|}
\hline
Models &IEEE $14$ Bus & IEEE $30$ Bus & IEEE $57$ Bus & IEEE $69$ Bus & IEEE $118$ Bus \\  \hline \hline
WLS                  &     $5.566$ ms  &    $0.0196$ s    &  $0.2917$ s         & $0.3867$ s         & $2.22$ s    \\ \hline
GM-Estimator                  &     -  &    $0.2133$ s    &  $1.23$ s         & -          & -    \\ \hline
MLP                  &     $0.0229$ ms &    $0.0294$ ms   &  $0.0512$  ms       & $0.0435$   ms       & $0.205$ ms           \\ \hline
CNN                  &     $0.095$ ms &    $0.1074$ ms    &  $0.211$ ms         & $0.196$ ms          & $0.53$ ms            \\ \hline
ProxlNet                  &     $0.0249$ ms &    $0.0347$ ms   &  $0.0535$ ms         & $0.0464$ ms         & $0.241$ ms   \\ \hline
Proposed                   &       $0.237$ ms   & $0.312$  ms        &  $0.597$ ms        & $0.591$  ms        & $2.03$  ms 
\\ 
Stacked ResNetD &&&&& \\ \hline

\end{tabular}
\label{tab:Runtime}
\end{table*}

\begin{figure}
    \includegraphics[scale=0.26]{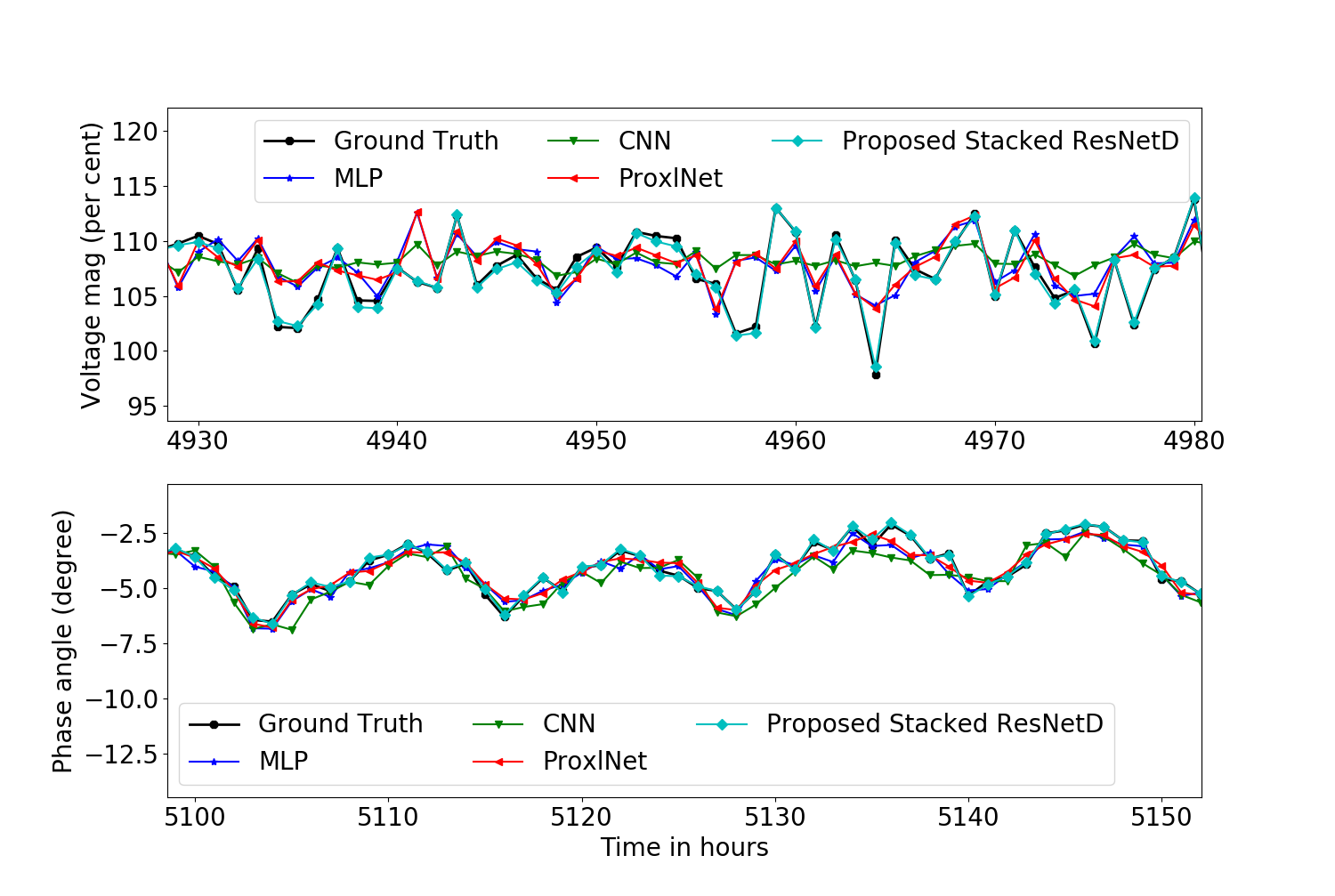}
    \caption{Prediction vs actual voltage magnitude and phase angles at bus $10$ of IEEE $14$ bus system}
    \label{fig:voltage_14_bus10}
\end{figure}

\begin{figure}
\centering
    \includegraphics[scale=0.26]{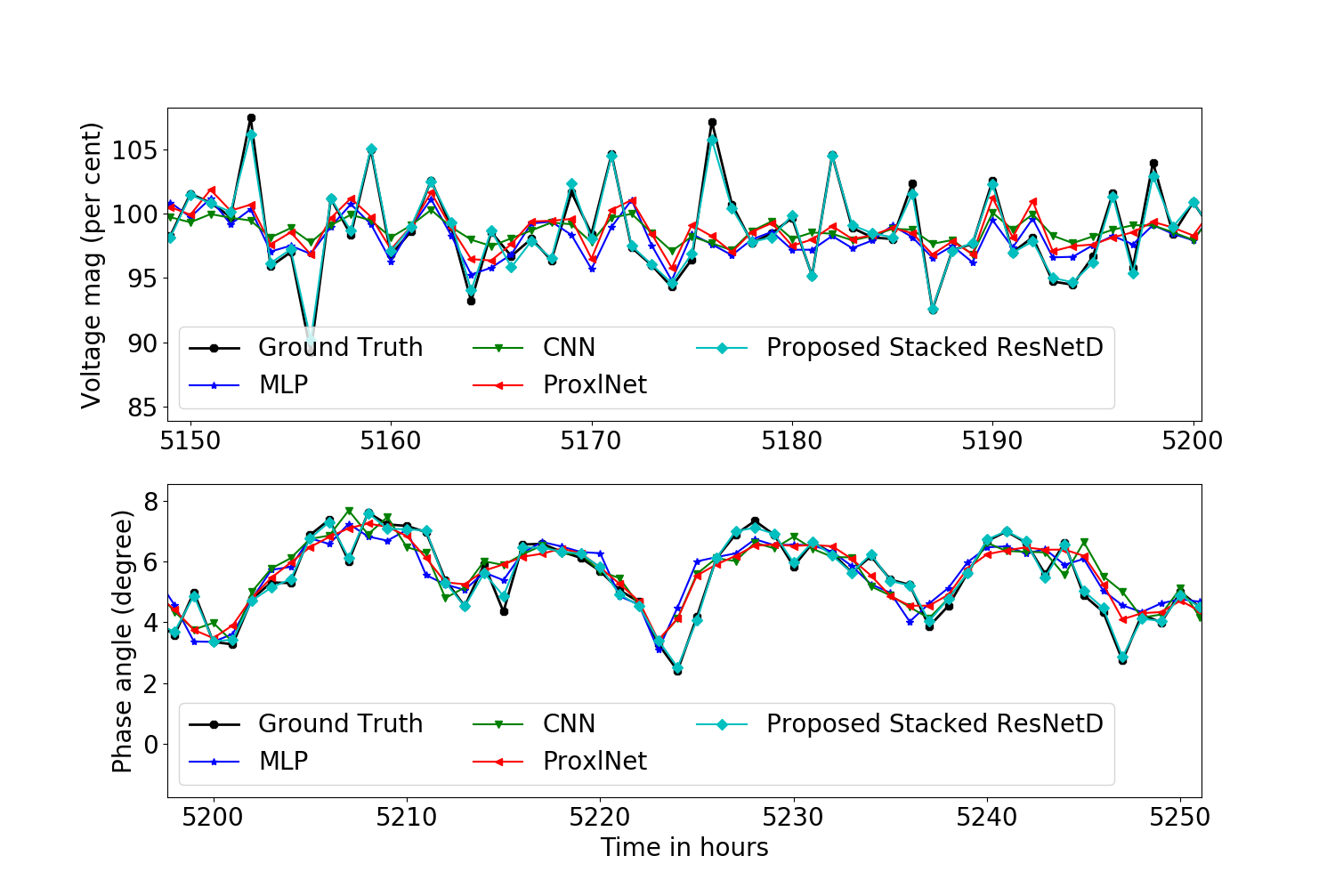}
    \caption{Voltage magnitude and phase prediction vs actual at bus $20$ of IEEE $30$ bus system}
    \label{fig:voltage_30_bus20}
\end{figure}

\begin{figure}
\centering
    \includegraphics[scale=0.26]{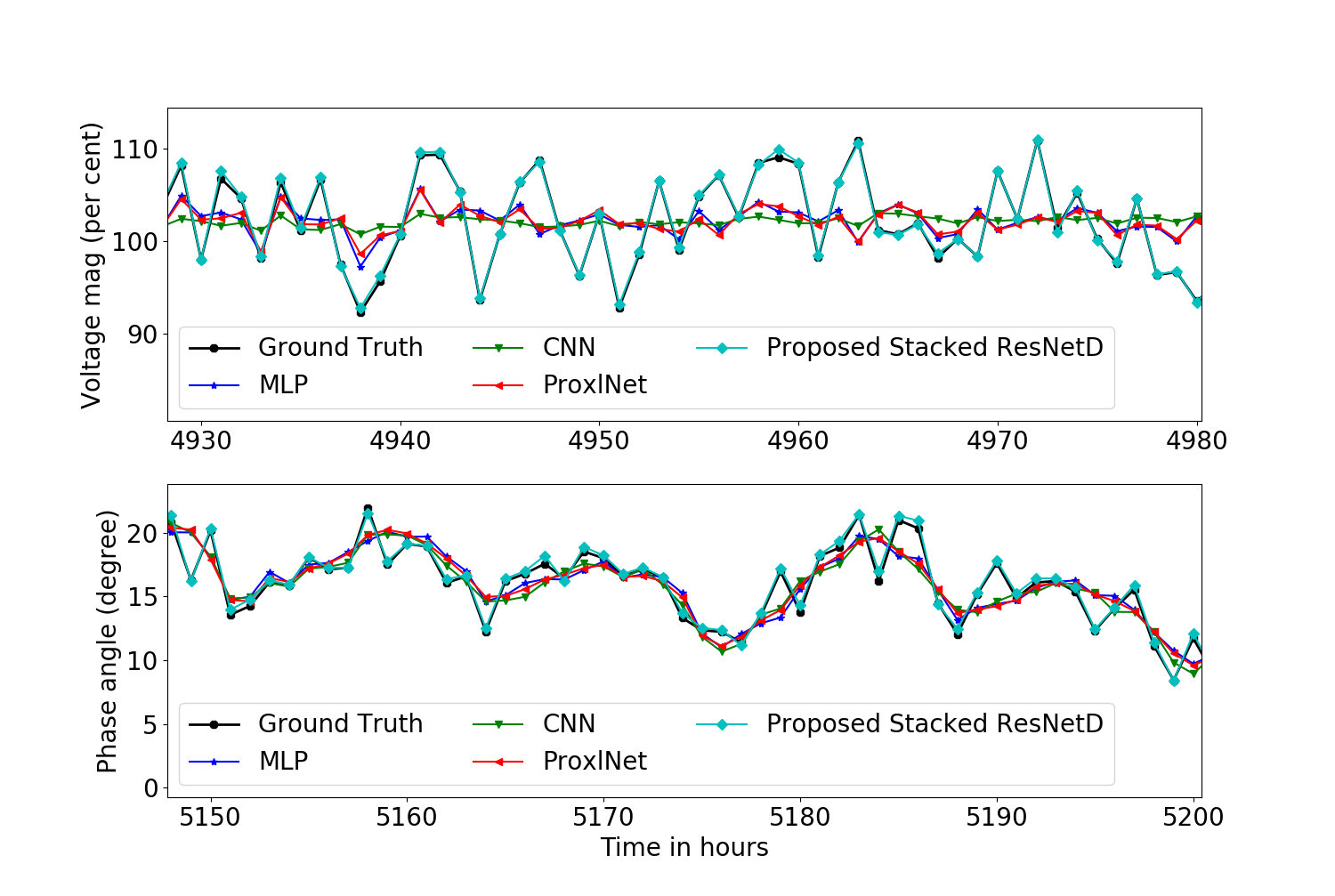}
    \caption{Prediction vs actual voltage magnitude and phase angles at bus $28$ of IEEE $57$ bus system}
    \label{fig:voltage_57_bus28}
\end{figure}

\begin{figure}
\centering
    \includegraphics[scale=0.26]{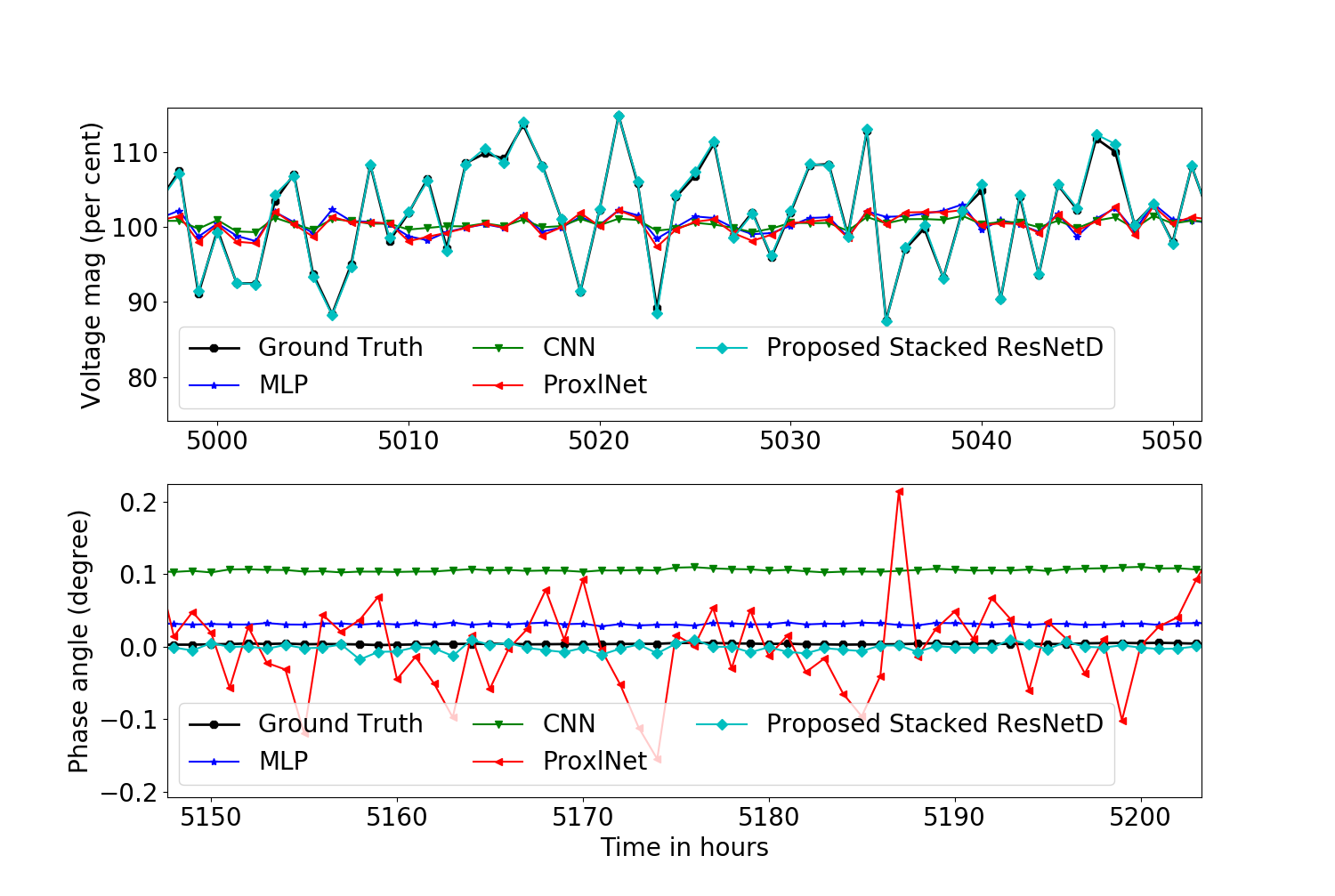}
    \caption{Prediction vs actual voltage magnitude and phase angles at bus $35$ of IEEE $69$ bus system}
    \label{fig:voltage_69_bus35}
\end{figure}

\begin{figure}
\centering
    \includegraphics[scale=0.26]{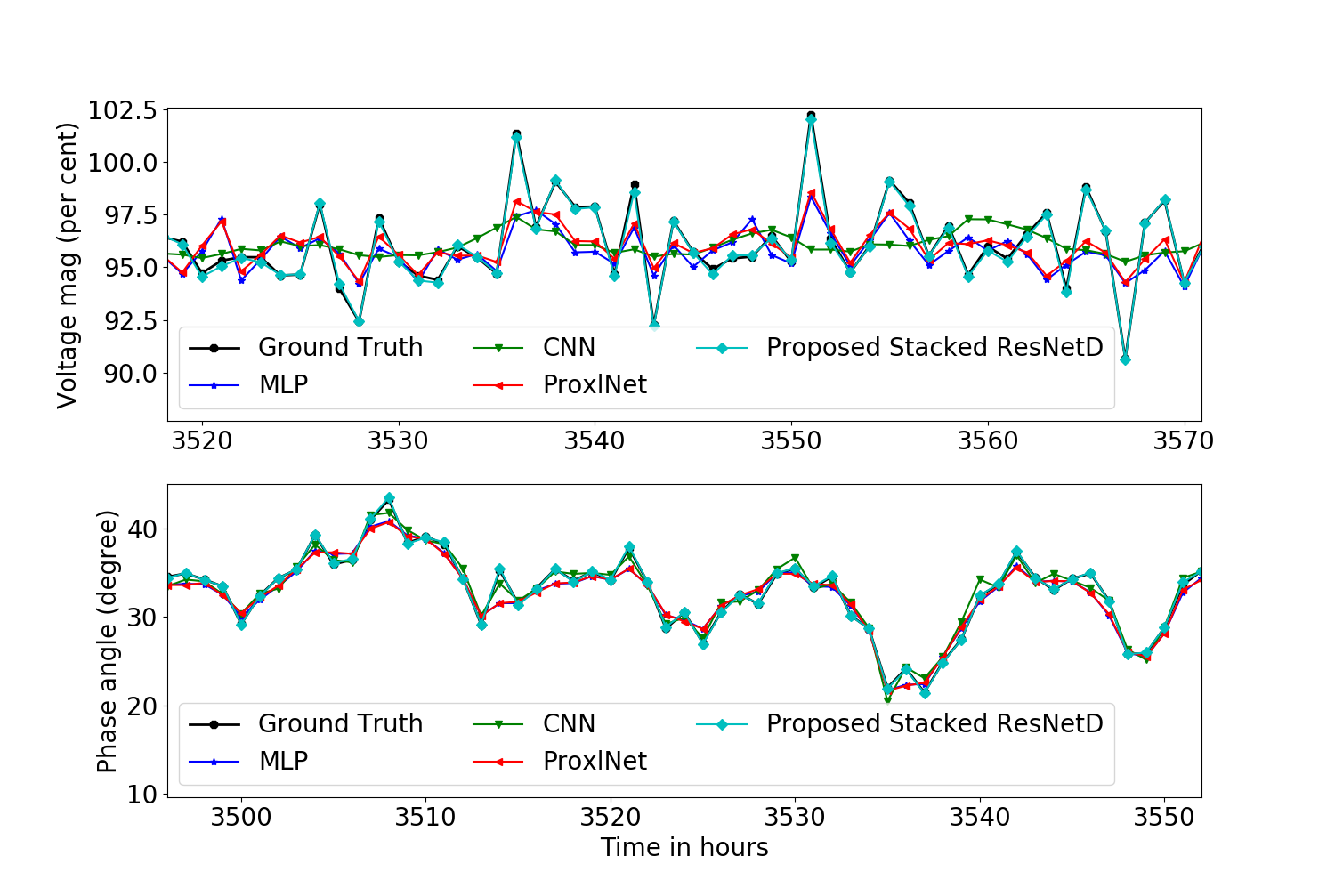}
    \caption{Prediction vs actual voltage magnitude and phase angles at bus $59$ of IEEE $118$ bus system}
    \label{fig:voltage_118_bus59}
\end{figure}

Figures~\ref{fig:voltage_14_bus10}, \ref{fig:voltage_30_bus20}, \ref{fig:voltage_57_bus28}, \ref{fig:voltage_69_bus35}, and \ref{fig:voltage_118_bus59} show the estimated voltage magnitudes and phase angles of the proposed Stacked ResNetD along with the MLP, CNN, and ProxlNet techniques at different buses of IEEE benchmark systems. These figures show that the states estimated by the proposed Stacked ResNetD are comparable to the actual states obtained by the WLS method. The states predicted by ResNetD are very close to the actual states. The states predicted by ResNetD as base-learners are linear approximations to actual states. Therefore, MLR as a meta-learner estimates the states with low bias.

\subsection{Performance of the Proposed Model with Non-Gaussian Noise}\label{sec:non_gaussian_noise}
We have used non-Gaussian noise to test the proposed approach on different distributions and noises. To test such cases, measurement errors are emulated randomly as follows. A random noise of size $0-3\%$ of original power flow results are generated and inserted in the measurement data. Separate random errors are generated for each measurement at every instance and added or subtracted from the original measurements. In this way, the error size and the disturbed measurements are changing every instance. Out of total of $m$ measurements, random errors are added in the first $50\%$ and subtracted from the remaining $50\%$ to make it more practical.

\begin{figure}
    \centering
    \includegraphics[scale=0.5]{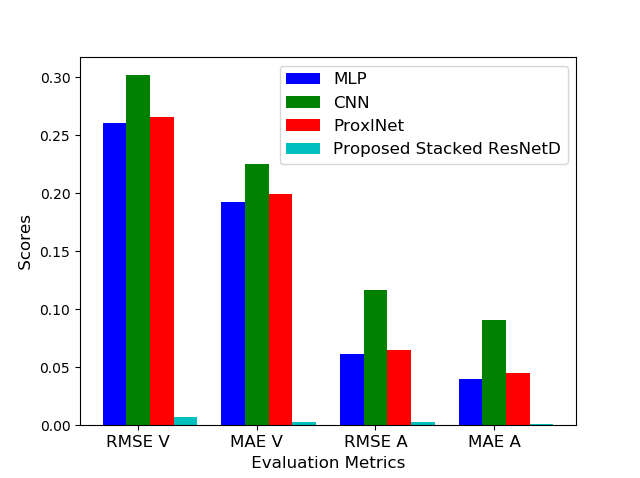}
    \caption{Performance of various models on IEEE 14 bus system with non-Gaussian measurement noise. RMSE V and MAE V are for voltage magnitudes and RMSE A and MAE A are for phase angles. Score in the vertical axis denotes voltage in percentage value for RMSE V and MAE V and phase angle in degree for RMSE A and MAE A.}
    \label{fig:non_Gauss_14}
\end{figure}

\begin{figure}
    \centering
    \includegraphics[scale=0.5]{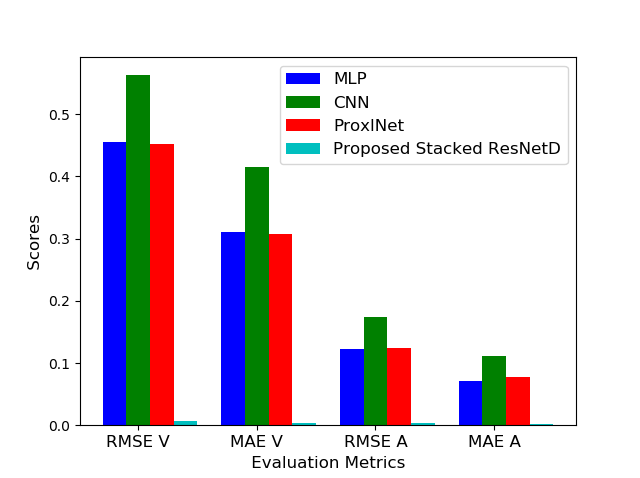}
    \caption{Performance of various models on IEEE 30 bus system with non-Gaussian measurement noise. RMSE V and MAE V are for voltage magnitudes and RMSE A and MAE A are for phase angles. Score in the vertical axis denotes voltage in percentage value for RMSE V and MAE V and phase angle in degree for RMSE A and MAE A.}
    \label{fig:non_Gauss_30}
\end{figure}

\begin{figure}
    \centering
    \includegraphics[scale=0.5]{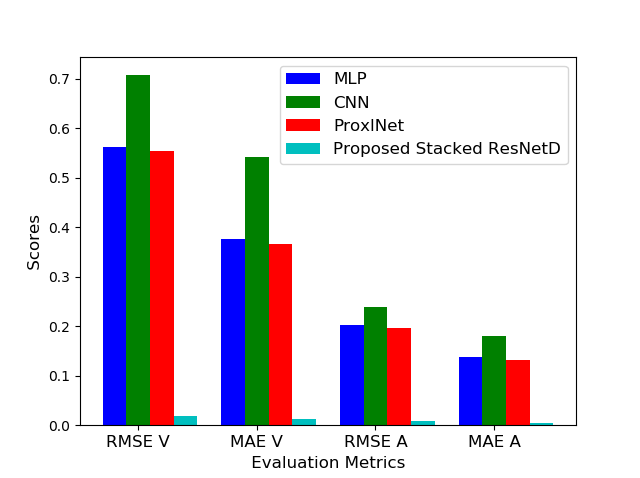}
    \caption{Performance of various models on IEEE 57 bus system with non-Gaussian measurement noise. RMSE V and MAE V are for voltage magnitudes and RMSE A and MAE A are for phase angles. Score in the vertical axis denotes voltage in percentage value for RMSE V and MAE V and phase angle in degree for RMSE A and MAE A.}
    \label{fig:non_Gauss_57}
\end{figure}

\begin{figure}
    \centering
    \includegraphics[scale=0.5]{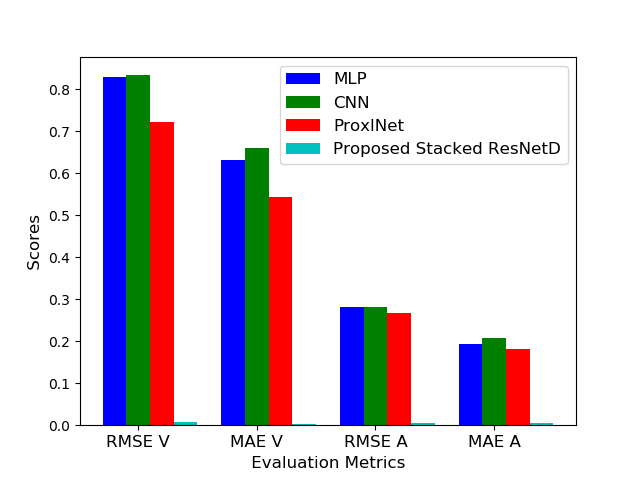}
    \caption{Performance of various models on IEEE 69 bus system with non-Gaussian measurement noise. RMSE V and MAE V are for voltage magnitudes and RMSE A and MAE A are for phase angles. Score in the vertical axis denotes voltage in percentage value for RMSE V and MAE V and phase angle in degree for RMSE A and MAE A.}
    \label{fig:non_Gauss_69}
\end{figure}

\begin{figure}
    \centering
    \includegraphics[scale=0.5]{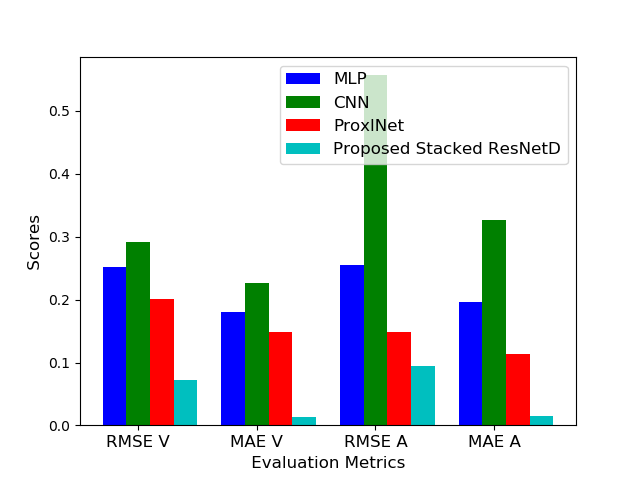}
    \caption{Performance of various models on IEEE 118 bus system with non-Gaussian measurement noise. RMSE V and MAE V are for voltage magnitudes and RMSE A and MAE A are for phase angles. Score in the vertical axis denotes voltage in percentage value for RMSE V and MAE V and phase angle in degree for RMSE A and MAE A.}
    \label{fig:non_Gauss_118}
\end{figure}

Figures~\ref{fig:non_Gauss_14}--~\ref{fig:non_Gauss_118} show the performance of MLP, CNN, ProxlNet, and the proposed Stacked ResNetD in terms of RMSE and MAE metrics. The results show that the proposed PSSE can estimate the states accurately even with non-Gaussian noise.

\subsection{Performance of Proposed PSSE with GM-Estimator}\label{GM_compare}
As the main purpose of the proposed work is to develop a machine learning model that can emulate the physical state estimator models appropriately, we have also tested it with generalised M-Estimator (GM-Estimator) to demonstrate the capability of the proposed method to emulate the GM-Estimator proposed in \cite{7742899, 496203}. The MATLAB code for GM-Estimator is obtained from \cite{7742899, 496203}. This case is tested only on the IEEE $30$- and $57$-node test systems because with the MATLAB code provided in \cite{7742899, 496203} for $69$  and $118$ bus system, GM-estimator state estimation is very time consuming and difficult to converge with time varying loads. There can be several ways to reduce the computational burden of the physical models which is outside the scope of the proposed work. To test the performance of proposed model with different number of input measurements, in this case the number of input measurements are different than that used in WLS based data generation: $80$ measurements are used for $30$ bus system, and $216$ measurements are used for $57$ bus system. In this case, the dataset are generated using a GM-Estimator. To incorporate Gaussian and non-Gaussian noise, datasets are generated with Gaussian noise and non-Gaussian noise and shuffled together and used for training and testing of the GM-Estimator. As there are $6$ base learners in proposed model, MLP, CNN, and ProxlNet are also run for $6$ times independently and the minimum of each run (prediction vary because of the stochastic nature of the deep learning models) is compared with the results of the proposed model. The per-unit values of voltage magnitudes are converted to percent values and the phase angles are converted from radian to degree for better visualisation.

For the $30$-bus system, RMSE for voltage magnitude estimation with MLP, CNN, ProxlNet, and the proposed Stacked ResNetD are $0.3815$, $0.3984$, $0.3518$, and $0.2551$, respectively. MAE for voltage magnitude estimation with MLP, CNN, ProxlNet, and the proposed Stacked ResNetD are $0.2663$, $0.2418$, $0.2510$, and $0.1353$, respectively. Similarly, for phase angle estimation, RMSE of MLP, CNN, ProxlNet, and the proposed Stacked ResNetD are $0.1689$, $0.2171$, $0.1603$, and $0.0741$, respectively. The phase angle estimation MAE of MLP, CNN, ProxlNet, and the proposed Stacked ResNetD are $0.0948$, $0.1103$, $0.0841$, and $0.025$, respectively. 

For the $57$-bus system, RMSE for voltage magnitude estimation of MLP, CNN, ProxlNet, and the proposed Stacked ResNetD are $1.2161$, $2.2138$, $1.022$, and $0.8897$, respectively. MAE for voltage magnitude estimation of MLP, CNN, ProxlNet, and the proposed Stacked ResNetD are $0.5714$, $0.8481$, $0.5545$, and $0.3805$, respectively. Similarly, for phase angle, RMSE of MLP, CNN, ProxlNet, and the proposed Stacked ResNetD are $0.4100$, $0.8304$, $0.3681$, and $0.2986$, respectively. The phase angle estimation MAE of MLP, CNN, ProxlNet, and the proposed Stacked ResNetD are $0.1638$, $0.2216$, $0.1644$, $0.0918$, respectively.


The performance of the proposed model is better than the other ML models because the proposed model with ResNetD as base learner can capture the non-linear relationship between the input measurements and the output states. The MLR as meta-learner further improves the results because of the approximate linear relationships between the output of the different ResNetD models and the actual states. The results show that the proposed model can accurately emulate the GM-Estimator.  

The run-time performance of machine learning models, WLS, and GM-Estimator is as shown in Table~\ref{tab:Runtime}.

\subsection{Results of MLR for State Forecasting}\label{perf_MLR_forecast}
When some of the measurements are missing during real-time operation, usually state forecasting is performed. In this section, MLR is compared with most common timer-series forecasting models such as CNN, LSTM, and hybrid CNN-LSTM for forecasting of the power system states. Each of these models is briefly discussed as follows.
\begin{itemize}
    \item CNN: The CNN used for the comparison consists sequentially of: two 1-D convolution layer with kernel size of $3$,  $64$ filters and  ReLU as activation function;  one 1-D max pooling layer with pool size of $2$;  one 1-D convolution layer with $128$ filters and kernel size of $3$; one 1-D global average pooling layer; one single dense layer with $50$ units and activation function of ReLU; and final dense layer with unit size equal to number of states to be forecasted.
    \item LSTM: LSTM has layered architecture. LSTM architecture used for the comparison consists sequentially of: three-layer of LSTM with ReLU activation function with $4n$, $2n$, and $2n$ units, respectively; and two dense layers with $2n$ number of units in each layer and ReLU as activation function.
    \item CNN-LSTM: The hybrid of CNN and LSTM consists of CNN networks followed by LSTM networks. CNN-LSTM used for the comparison consists of two layers of 1-D convolution layer with $64$ filters, kernel size of $3$ in each, and ReLU as activation function; one 1-D max pooling layer with pool size of $2$; two LSTM layers with $2n$ number of units with ReLU activation function; and three dense layers with $2n$ units with ReLU activation function in each layer. 
\end{itemize}
Adam is used as an optimiser and mean absolute error is taken as a loss function for all of the models. All of the CNN, LSTM, and CNN-LSTM models are run $6$ times independently with batch size of $32$ and $200$ epochs, and the minimum values of metrics of all runs are taken for the purpose of comparison.

Out of the available historical data, $40\%$ are used for training and the remaining $60\%$ are used for testing. For this time-series forecasting, the last $24$ step time series data of states are utilised to forecast current hour's state. Although a state is forecasted only for one step (one hour), the proposed approach can be used to forecast the states for multiple steps (two or more hours) with a little modification.

Table~\ref{tab:comparisonforecastingvoltage} provides comparisons between CNN, LSTM, CNN-LSTM, and MLR models in terms of RMSE and MAE metrics forecast of voltage magnitudes for IEEE $14$, $30$, $57$, $69$, and $118$ benchmark systems on test datasets. It can be seen from the table that the performance of MLR is remarkably better than the other models for all of the tested systems.

Table \ref{tab:comparisonforecastingphase} presents a comparison between forecast of phase angles of MLR against LSTM, CNN-LSTM, and CNN on test data-set of IEEE $14$, $30$, $57$, $69$, and $118$ bus benchmark systems in terms of RMSE and MAE metrics. It can be seen that MLR outperforms all other models for all of the studied IEEE benchmark systems.

\begin{table*}[]
\caption{Comparison of voltage magnitude forecast of CNN, LSTM, CNN-LSTM, and MLR models in terms of RMSE and MAE metrics for standard IEEE $14$, $30$, $57$, $69$, and $118$ benchmark systems}
\centering
\begin{tabular}{|l|l|l|l|l|l|l|l|l|l|l|}
\hline
\multirow{2}{*}{Models} & \multicolumn{2}{l|}{IEEE $14$ Bus} & \multicolumn{2}{l|}{IEEE $30$ Bus} & \multicolumn{2}{l|}{IEEE $57$ Bus} & \multicolumn{2}{l|}{IEEE $69$ Bus} & \multicolumn{2}{l|}{IEEE $118$ Bus} \\ \cline{2-11} 
& RMSE          &     MAE      & RMSE           &  MAE          & RMSE           &    MAE & RMSE           &  MAE          & RMSE           &    MAE      \\ \hline

LSTM                   &     $0.4174$  &    $0.2927$    &  $0.3351$         & $0.2397$          & $0.9008$          & $0.5772$ & $0.5312$ & $0.3554$& $0.5271$& $0.3751$          \\ \hline

CNN-LSTM                   &     $0.4233$  &    $0.3029$    &  $0.3349$         & $0.2382$          & $0.9028$          & $0.6159$ & $0.5278$ & $0.3715$& $0.5108$& $0.3421$          \\ \hline

CNN                   &     $0.4218$  &    $0.3233$    &  $0.3640$         & $0.2814$          & $0.6700$          & $0.5799$ & $0.4946$ & $0.3444$& $0.4868$& $0.3305$          \\ \hline

Proposed                    &     $\bold{0.1241}$  &    $\bold{0.0845}$    &  $\bold{0.1509}$         & $\bold{0.1044}$          & $\bold{0.2115}$          & $\bold{0.1417}$ & $\bold{0.2529}$ & $\bold{0.1955}$& $\bold{0.1919}$& $\bold{0.1420}$      \\ MLR &&&&&&&&&&    \\ \hline
\end{tabular}
\label{tab:comparisonforecastingvoltage}
\end{table*}

\begin{table*}[]
\caption{Comparison of phase angles forecast of CNN, LSTM, CNN-LSTM, and MLR models in terms of RMSE and MAE metrics for standard IEEE $14$, $30$, $57$, $69$, and $118$ benchmark systems}
\centering
\begin{tabular}{|l|l|l|l|l|l|l|l|l|l|l|}
\hline
\multirow{2}{*}{Models} & \multicolumn{2}{l|}{IEEE $14$ Bus} & \multicolumn{2}{l|}{IEEE $30$ Bus} & \multicolumn{2}{l|}{IEEE $57$ Bus} & \multicolumn{2}{l|}{IEEE $69$ Bus} & \multicolumn{2}{l|}{IEEE $118$ Bus} \\ \cline{2-11} 
& RMSE          &     MAE      & RMSE           &  MAE          & RMSE           &    MAE & RMSE           &  MAE          & RMSE           &    MAE      \\ \hline

LSTM                   &     $0.0337$  &    $0.0263$    &  $0.0399$         & $0.0318$          & $0.0853$          & $0.0647$ & $0.0601$ & $0.0516$& $0.2182$& $0.1594$          \\ \hline

CNN-LSTM                   &     $0.0399$  &    $0.0325$    &  $0.0526$         & $0.0392$          & $0.1108$          & $0.0902$ & $0.0602$ & $0.0526$& $0.2037$& $0.1525$          \\ \hline

CNN                   &     $0.0686$  &    $0.0575$    &  $0.0633$         & $0.0492$          & $0.0864$          & $0.0707$ & $0.0659$ & $0.0562$& $0.2325$& $0.1730$          \\ \hline

Proposed                    &     $\bold{0.0037}$  &    $\bold{0.0025}$    &  $\bold{0.0036}$         & $\bold{0.0025}$          & $\bold{0.0094}$          & $\bold{0.0067}$ & $\bold{0.0015}$ & $\bold{0.0010}$& $\bold{0.0460}$& $\bold{0.0343}$   \\ MLR &&&&&&&&&&       \\ \hline
\end{tabular}
\label{tab:comparisonforecastingphase}
\end{table*}

\begin{figure}
\centering
    \includegraphics[scale=0.26]{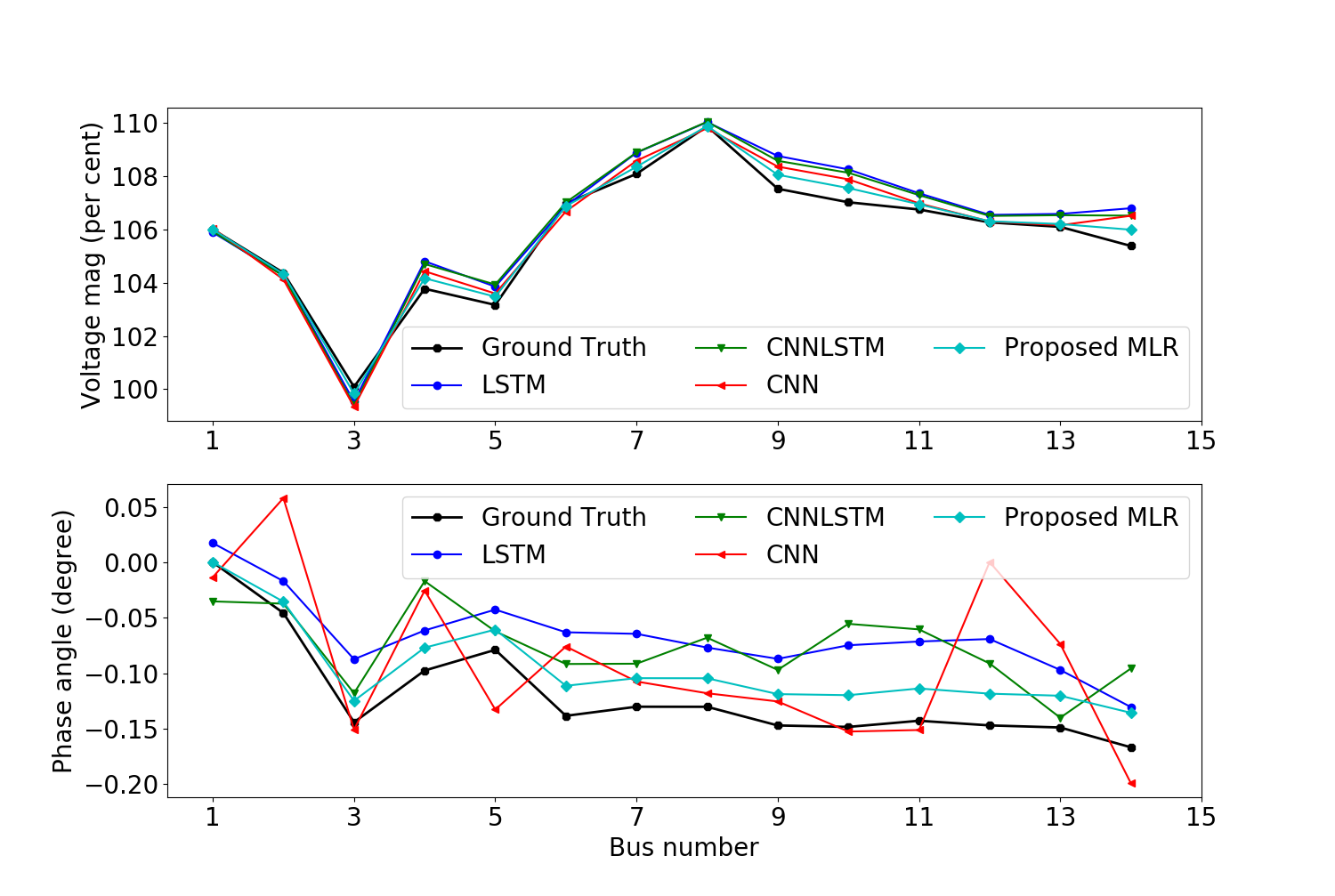}
    \caption{Prediction vs actual voltage magnitude and phase angles at all buses of  IEEE $14$ bus system at instant $500$}
    \label{fig:voltage_forecast_14}
\end{figure}

\begin{figure}
\centering
    \includegraphics[scale=0.26]{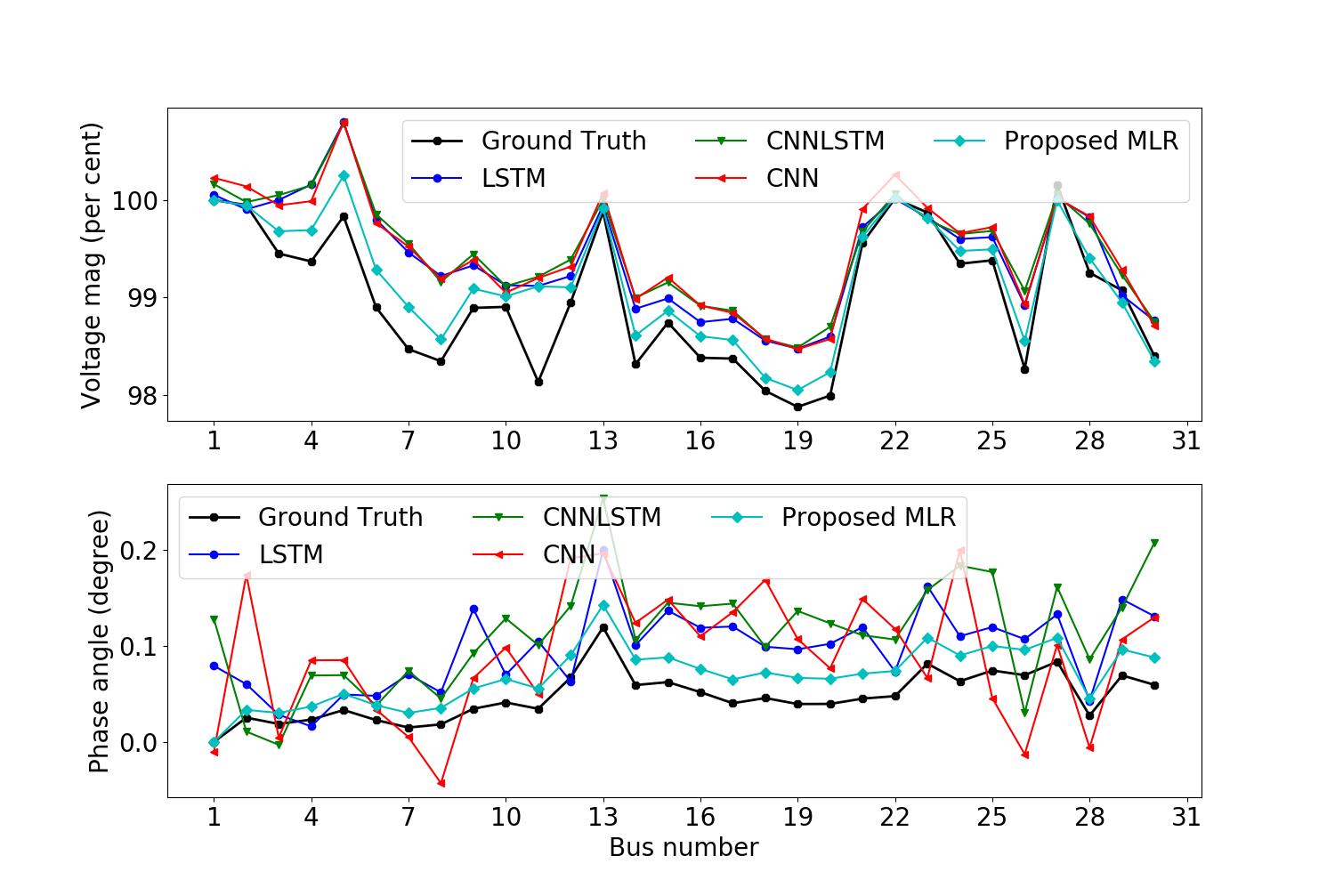}
    \caption{Prediction vs actual voltage magnitude and phase angles at all buses of  IEEE $30$ bus system at instant $500$}
    \label{fig:voltage_forecast_30}
\end{figure}

\begin{figure}
\centering
    \includegraphics[scale=0.26]{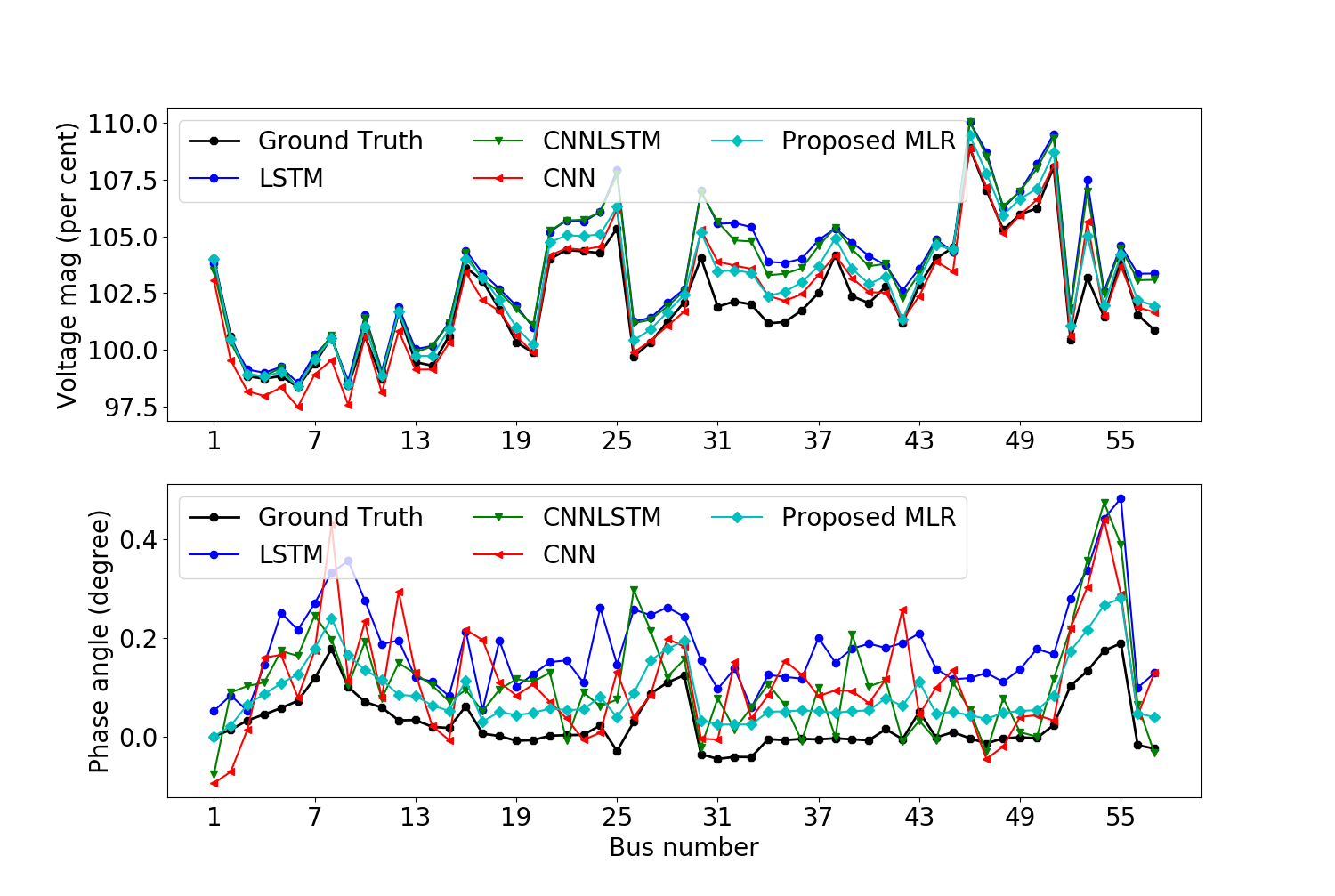}
    \caption{Prediction vs actual voltage magnitude and phase angles at all buses of  IEEE $57$ bus system at instant $500$}
    \label{fig:voltage_forecast_57}
\end{figure}

\begin{figure}
\centering
    \includegraphics[scale=0.26]{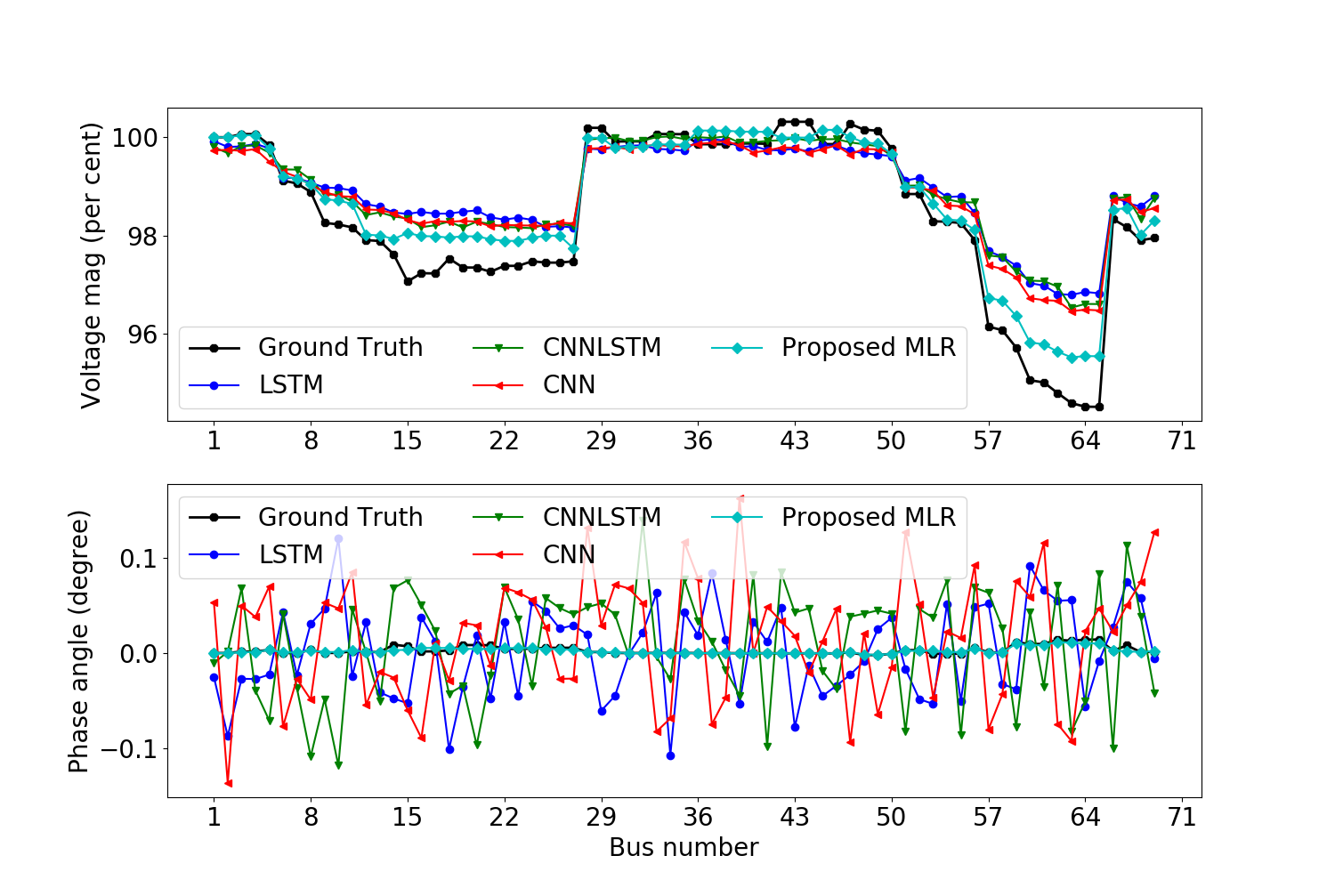}
    \caption{Prediction vs actual voltage magnitude and phase angles at all buses of  IEEE $69$ bus system at instant $500$}
    \label{fig:voltage_forecast_69}
\end{figure}

\begin{figure}
\centering
    \includegraphics[scale=0.27]{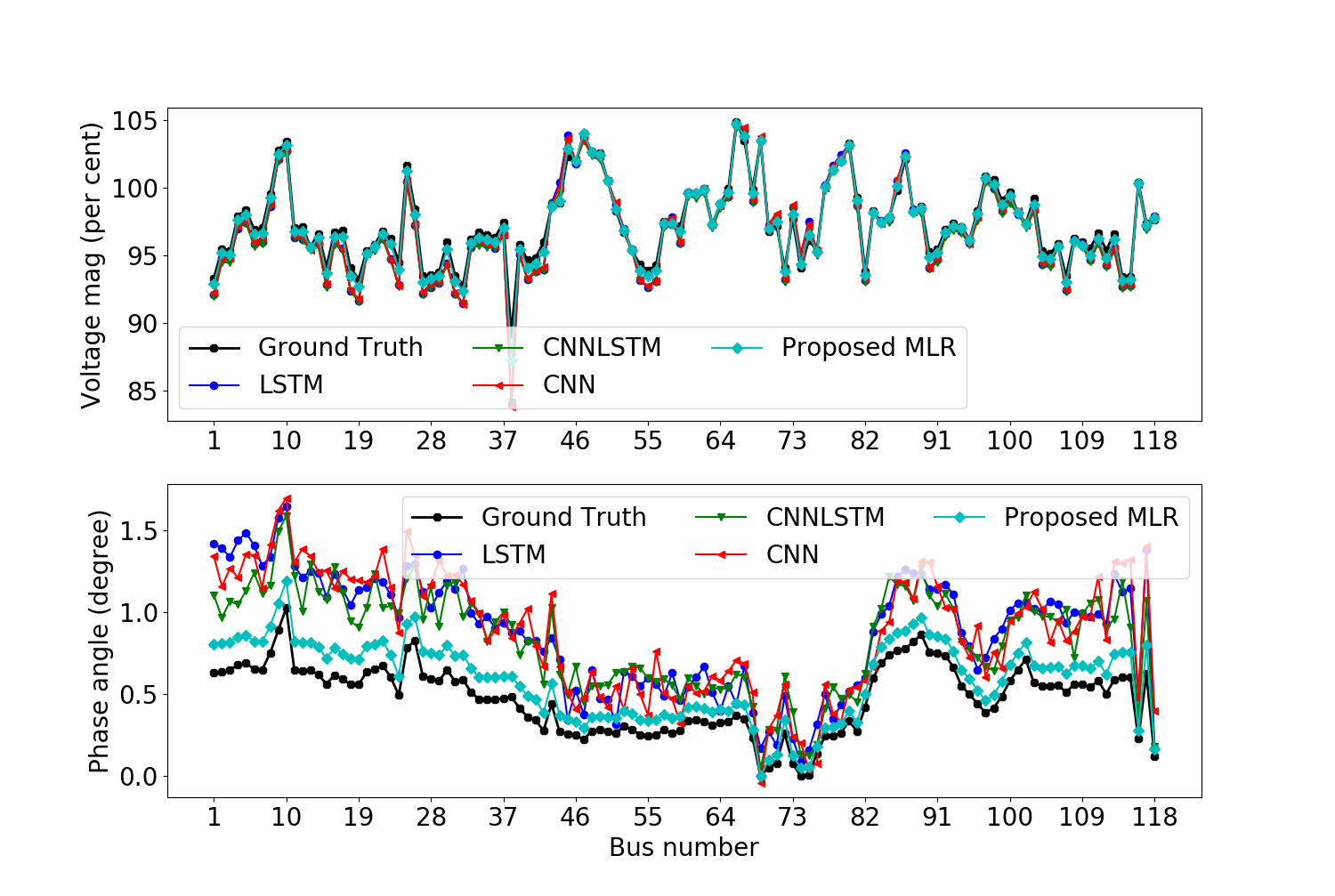}
    \caption{Prediction vs actual voltage magnitude and phase angles at all buses of  IEEE $118$ bus system at instant $500$}
    \label{fig:voltage_forecast_118}
\end{figure}

Figures~\ref{fig:voltage_forecast_14}, \ref{fig:voltage_forecast_30}, \ref{fig:voltage_forecast_57}, \ref{fig:voltage_forecast_69}, and \ref{fig:voltage_forecast_118} show the comparison between CNN, LSTM, CNN-LSTM, and MLR to forecast the power system states. These figures show the competitive performance of MLR for state forecasting. 

MLR is mapping the historical states closer to actual states than any other compared deep learning models. This could be due to the existence of an approximately linear relationship between historical power system states.

Although the proposed state forecasting approach can forecast the current states when all sets of measurements are available, the states thus obtained are not as close as estimated states using the proposed PSSE with the current measurements (the comparison results are not provided for obviousness and simplicity of expositions, if interested it can be verified with provided source code).  Therefore, the state forecasting should only be used at the instant of missing measurements. 

\section{Conclusions}\label{conclusion}
This paper has proposed a data-driven real-time PSSE using a deep ensemble learning method. The proposed deep ensemble learning setup was formed by stacking several parallel ResNetD as base-learners and multivariate-linear regression as meta-learner.  In this work, historical measurements and states were utilised to train the proposed model for the estimation of power system states (voltage magnitudes and phase angles). The trained model was utilised to predict the states of the power system in real time using real-time measurements. The data-driven PSSE assumes the availability of a complete set of measurements; however, some of the real-time measurements may be missing leading to failure in estimating the states. To deal with missing measurements, this paper adopted multivariate-linear regression to forecast the missing states at any instant using historical states. Several case studies were performed in various IEEE benchmark systems. Case studies showed that the proposed approach outperformed various machine learning techniques.

\section*{Declaration of Competing Interest}
None.
\section*{Acknowledgement}
This work was supported by the U.S. National Science Foundation (NSF) under Grant NSF 1847578.
\appendix
 \section{Supplementary material}
 Source code developed for this project are publicly available at GithuB.

\bibliographystyle{elsarticle-num}
\bibliography{References}








\end{document}